\documentclass{article} \usepackage{iclr_preprint,times}
\iclrfinalcopy

\usepackage{amsmath,amsfonts,bm}

\def\eqref#1{equation~\ref{#1}}

\def\1{\bm{1}}

\DeclareMathAlphabet{\mathsfit}{\encodingdefault}{\sfdefault}{m}{sl}
\SetMathAlphabet{\mathsfit}{bold}{\encodingdefault}{\sfdefault}{bx}{n}

\usepackage{hyperref}
\usepackage{url}

\usepackage[utf8]{inputenc} \usepackage[T1]{fontenc}    \usepackage{hyperref}       \usepackage{url}            \usepackage{booktabs}       \usepackage{amsfonts}       \usepackage{nicefrac}       \usepackage{microtype}      \usepackage{xcolor}         \usepackage{xspace}
\usepackage{amssymb}

\usepackage{amsmath}
\usepackage{cleveref}
\usepackage{graphicx}
\usepackage{wrapfig}
\usepackage{subcaption}
\usepackage{array}

\newcommand{\Rb}{\mathbb{R}}
\newcommand{\kl}{\textsc{kl}}
\newcommand{\argmax}[1]{\underset{#1}{\mathrm{arg\,max}}}
\newcommand{\argmin}[1]{\underset{#1}{\mathrm{arg\,min}}}
\newcommand{\Eb}{\mathbb{E}}
\newcommand{\Xc}{\mathcal{X}}

\newcommand{\Hc}{\mathcal{H}}

\newcommand{\Uc}{\mathcal{U}}

\newcommand{\hla}{HLA-A*02:01\xspace}

\title{Continuous Variational Synthesis}

\author{Alan N. Amin$^\star$, Mattia G. Gollub$^\star$, Andrei Slabodkin, Elizabeth B. Wood$^\dagger$, Eli N. Weinstein$^\dagger$ \\
JURA Bio, Inc.\\
Boston, MA 02125, USA 
}

\begin{document}

\maketitle

\begin{abstract}
    Biological machine learning was long bottlenecked by the ability to synthesize designed DNA. 
    Variational synthesis models control chemical reactions to physically manufacture quadrillions of designed sequences in DNA. However, training these generative models is challenging: constraints on chemical synthesis can force many parameters into a discrete space, limiting the ability to pre-train and fine-tune.
    In this article we train ``free'' variational synthesis models using stochastic gradient descent in continuous space, and then discretize  with post-training quantization to impose hardware and wetware constraints.
    This enables variational synthesis models to satisfy stringent reward criteria, while still synthesizing diverse designs, achieving a strictly dominating quality-diversity Pareto frontier.
    We demonstrate by training variational synthesis models of enzymes, peptides, antibody CDRH3s, and regulatory DNA elements. \textit{In silico} performance is maintained \textit{in vitro}.
\end{abstract}

Advances in biological machine learning have provided a rich set of tools for designing DNA, RNA and proteins \citep{Koh2025-rv,Gosai2024-oe}. 
Our ability to learn about these designs experimentally depends on our ability to make them in the lab.
Traditionally, designs are synthesized individually and deterministically. But this approach is limited in its scalability, bottlenecking our ability to learn about the activity of biological sequences.

Recent work constructs petascale DNA libraries of ${\sim}10^{16}$ designs using stochastic chemical synthesis controlled by generative models \citep{Weinstein2022-sw,Weinstein2026-eg}.
This \textit{variational synthesis} approach requires training generative models that satisfy underlying synthetic constraints.
However, these constraints are often discrete and high-dimensional, and therefore difficult to search over: many parameters may only take values from a finite catalog.
As a result, training runs often get stuck in local minima, and many fine-tuning and reinforcement learning methods are unavailable.

Motivated by advances in training language models, we set out to develop improved methods for training variational synthesis models for DNA, RNA and proteins. 
Our key idea is to train variational synthesis models in a ``free'' continuous space, and then discretize them through a post-training quantization procedure.
In the continuous space, we can optimize using stochastic gradient descent methods. We derive custom variance reduction methods that exploit the model structure to further accelerate training.  
Post-training quantization allows us to generate samples from the model on diverse hardware and wetware platforms. 
We adjust the model's parameters to meet the synthesizer's constraints, then run stochastic synthesis to manufacture designs at petascale.

Overall, the approach substantially improves forward KL pre-training of variational synthesis models, and enables reverse KL fine-tuning.
We demonstrate in silico by designing libraries of enzymes that fold into a specific structure; peptides predicted to bind an HLA; scFvs predicted to target intracellular antigens based on experimental feedback; and promoters predicted to drive T cell-specific expression. On each, we substantially advance the quality-diversity Pareto frontier.
We then verify by sequencing that the quality and diversity is maintained \textit{in vitro}.

\subsection{Background: Variational Synthesis}

The standard protocol for synthesizing designs from a generative model is to first sample designs computationally, then synthesize those designs individually.
In variational synthesis, sampling happens during synthesis, using generative model-controlled stochastic chemical reactions.

\paragraph{Synthesis model.} 
DNA is commonly synthesized by solid-phase oligonucleotide synthesis, which builds DNA strands by sequentially adding nucleotides \citep{Gait1978-zm}.
If we add a mixture of nucleotides at the same time, each growing DNA molecule will randomly encounter a different nucleotide, according to its concentration in the mixture.
If we add a mixture $\theta_\ell\in\Delta^4$ at step $\ell$, then we can describe the output of the synthesis procedure with a parameterized probability distribution
$q_\theta(X)=\prod_{\ell=1}^L\theta_{\ell, X_\ell}$
where $X_{\ell}$ is the $\ell$-th position of sequence $X$, which has total length $L$.
If we run the synthesis $M$ times in different wells with different parameters, and mix the results in equal concentrations, we get:
\begin{equation} \label{eqn:vs-model}
    q_\theta(X)  = \frac 1 M\sum_{m=1}^M \prod_{\ell=1}^{L} \theta^m_{\ell, X_\ell}.
\end{equation}
This model can describe DNA or RNA sequences or, after translation, the distribution of proteins the synthesized DNA encodes \citep{Weinstein2022-sw}.
In this paper we fix the weights to $1/M$ and length per well to $L$ for simplicity, but in general these parameters can also be controlled by adjusting the relative concentration of the products and the number of reaction steps.

\paragraph{Constraints.} There are often constraints on $\theta$ imposed by chemistry and chemical engineering.
Synthesizers can restrict $\theta_\ell^m$ to take on values from a finite catalog, since mixtures must be pulled from a finite set of flasks. 
One common restriction is the $2^4-1=15$ equal-weight nucleotide mixtures $\mathcal U=\{(1, 0, 0, 0), (0, 1, 0, 0), \dots, (\frac 1 2, \frac 1 2, 0, 0), (\frac 1 2, 0, \frac 1 2, 0), \dots, (\frac 1 3, \frac 1 3, \frac 1 3, 0), \dots\}.$
These can sometimes be supplemented by $K$ pre-set nucleotide mixes, $\Uc_\psi = \{(1, 0, 0, 0), \ldots, (\frac 1 2, \frac 1 2, 0, 0), \ldots, \psi_1, \ldots, \psi_K\}$, where $\psi_k \in \Delta^4$ is shared across all $\ell,m$. 
In either case, $\theta_\ell^m$ cannot be an arbitrary vector in $\Delta^4$, but instead must satisfy the constraint: $\theta_\ell^m \in  \mathcal{U} \subset \Delta^4$ for all $\ell, m$.

\paragraph{Training.} Variational synthesis models have been trained via a \textit{forward KL} objective, maximizing data log likelihood:
\begin{equation} \label{eqn:fwd-kl}
    \theta_\star = \argmax{\theta}\, \frac{1}{n} \sum_{i=1}^n \log q_\theta(X^{(i)}).
\end{equation}
Optimizing this objective corresponds to approximately minimizing $\textsc{kl}(p \| q_\theta)$, where $p$ is the distribution of the data.
Previous work used online expectation maximization (EM) algorithms: every M-step optimized each $\theta^m_\ell$ to the best choice in $\mathcal U$ \citep{Weinstein2022-sw,Cappe2008-gs}. But EM is prone to local maxima.
Concurrent work proposes \textit{policy gradient library design} (PGLD) which optimizes an expected reward under the distribution $q_\theta(x)$ \citep{Sussex2026-am}.
PGLD handles constraints by parameterizing a ``meta-distribution'' $q_\phi(\theta)$ and optimizing $\phi$ in $\mathbb E_{q_\phi(\theta)}\mathrm{loss}(\theta)$.
We evaluate both methods in depth in \Cref{sec:results}.

\paragraph{Synthesis.}
In the lab, we run stochastic synthesis with the learned parameters $\theta_\star$. Then each synthesized molecule is an independent sample from the trained model, $X \sim q_{\theta_\star}(x)$. The total number of samples is the yield of the synthesis.
For standard oligosynthesis, this yield can be trillions or quadrillions of molecules, meaning variational synthesis can produce DNA designs at petascale \citep{Weinstein2026-vs}.

\subsection{Related Work} \label{sec:related}

We build on variational synthesis \citep{Weinstein2022-sw,Weinstein2026-vs}.
Previous work has also developed computational methods for controlling stochastic DNA synthesis, though without optimizing a probabilistic model of the library \citep{jacobs_swiftlib_2015,mena_automated_2005, shimko_decode_2020,parker_optimization_2011}. 
\citet{Zhu2024-ad} consider an objective related to the reverse KL, but only study synthesis with $M =1$ reaction well.
Most closely related is concurrent work by \citet{Sussex2026-am}, who study objectives related to the reverse KL and use complex synthesis models ($M > 1$).

Our approach can be seen as variational inference with a mixture model, a successful technique for approximate Bayesian computation at scale \citep{Lin2019-wu,Lin2020-wu,Wilson2022-ak,Arenz2022-gx,Lambert2022-fo,Petit-Talamon2025-gl}.
We use the model to control experiments rather than just to approximate a posterior, so face distinct constraints.
This perspective aligns with other efforts to control generative models using probabilistic inference \citep{Levine2018-yf,Korbak2022-hd,Wu2023-yj,Zhao2024-kt}.

\section{Method}

We set out to develop improved methods for training variational synthesis models, motivated by advances in training language models.
\begin{enumerate}
    \item In addition to \textit{pre-training} with a forward KL objective, we \textit{fine-tune} with a reverse KL objective, to optimize explicit reward models such as sequence-activity predictors. \item We optimize both objectives with a unified \textit{gradient-based training} approach, and introduce model-specific strategies for reducing gradient variance.
\item  We develop a \textit{quantization} technique to impose chemical and synthesis constraints, which preserves the quality and diversity of the underlying model in practice.
\end{enumerate}
We refer to the combined improvements as continuous variational synthesis (cVS).

\subsection{Fine-tuning}
In addition to the forward KL pre-training objective (\Cref{eqn:fwd-kl}), we consider a reverse KL fine-tuning objective,
\begin{equation} \label{eqn:rev-kl}
    \theta_\star = \argmax{\theta}\, \Eb_{q_\theta}[r(X)] - \kl(q_\theta \| \pi)
\end{equation}
where $r: \Xc \to \Rb$ is a reward model that assigns a scalar score to a sequence $x$, and $\pi(x)$ is a prior distribution, such as a generative sequence model pre-trained on human or evolutionary data. Optimizing this objective corresponds to minimizing the reverse KL divergence $\textsc{kl}(q_\theta \| \tilde p)$ to the prior tilted by the reward, $\tilde p(x) \propto \pi(x) \exp(r(x))$.
This objective is widely used in language model fine-tuning, where it is also referred to as regularized reinforcement learning \citep{Korbak2022-hd}.

The forward KL (\Cref{eqn:fwd-kl}) prioritizes coverage: the library distribution should assign a high likelihood to every sequence in the training data, so no areas of sequence space are missing (it is \textit{mode-covering}).
The reverse KL (\Cref{eqn:rev-kl}) prioritizes quality: every sequence from the library distribution should have a high reward, even if this means not making sequences from some areas of sequence space (it is \textit{zero-avoiding}). 
This tradeoff can be especially preferable in later exploitation phases of an engineering campaign.

\subsection{Gradient-Based Training of Free VS}
To train variational synthesis models, we first relax the synthesis constraints to train a ``free'' model.
We let each $\theta^m_\ell=\mathrm{softmax}(\phi_{\ell}^m)$ where $\phi_{\ell}^m\in \mathbb R^4$ if $X$ is DNA or RNA and $\phi_{\ell}^m\in \mathbb R^{21}$ if $X$ is a protein sequence. Now $\theta_\ell^m$ can take any value on the simplex.
The gradient becomes,
\begin{align}
\frac{1}{n} \sum_{i=1}^n \nabla_{\phi} \log q_\theta(X) & \quad \quad \quad \quad  \textit{forward\, KL} \\
    \nabla_{\phi} \Eb_{q_{\theta}}[r(X)] - \nabla_{\phi} \kl(q_{\theta} \| \pi) & \quad \quad \quad \quad \textit{reverse\, KL}.
\end{align}
We then optimize using Adam \citep{Kingma2015-ej}.
For the reverse KL, computing the gradient with respect to an expectation over the model is nontrivial.
Naive REINFORCE estimators have very high variance \citep{Williams1992-xb}. 
To reduce variance we start by decomposing the objective into 
\begin{equation}
    \Eb_{q_\theta}[r(X)] + \Eb_{q_\theta}[\log \pi(X)] + \Hc(q_\theta)
\end{equation}
where $\Hc(q_\theta)\triangleq-\Eb_{q_\theta}[\log q_\theta(X)]$ denotes the entropy. We develop partially analytic approximations to the last two terms, which exploit the model structure.
Additional training details are in \Cref{apx:training-techniques}.

\subsubsection{Analytic entropy} 
To reduce variance, we rewrite the entropy $\Hc(q_\theta)$ using the mixture model structure of the variational synthesis model. 
We introduce $z$, a latent variable indicating the mixture model component, i.e. which reaction well $m$ synthesizes $x$. Let $q(X \mid z=m) = \prod_{\ell=1}^{L} \theta_{\ell, X_\ell}^m$ denote the probability of a sequence in well $m$, and let $r(z=m \mid X) =  q_\theta(X | m)/(Mq_\theta(X))$ denote the \textit{responsibility} of well $m$, i.e. the posterior likelihood that $x$ comes from well $m$.
With some algebra, we have
\begin{equation}
    \Hc(q_\theta) =\frac 1 M \sum_{m=1}^M  \Hc(q(X \mid m)) + \log(M) - \Eb_{q_\theta}[\Hc(r(z \mid X))]
\end{equation}
The first term favors high diversity at each position in each well, while the last  
prefers to minimize sequence overlap between the wells. 
The gradient of the first two terms can be calculated analytically, eliminating any variance; only the last requires a REINFORCE estimate. 

\subsubsection{Prior Rao-Blackwellization} 
To further reduce gradient variance, we next examine the contribution of the prior to the reverse KL objective, $\Eb_{q_\theta} [\log \pi(x)] $. 
We are specifically interested in priors $\pi(x)$ specified by autoregressive models, which include most protein and genomic language models.
Then,
\begin{equation} \label{eqn:ar-rb}
    \Eb_{q_\theta} [\log \pi(X)] = \sum_{\ell=1}^{L}\mathbb E_{q_\theta(X_{1:\ell-1})}\mathbb E_{q_\theta(X_{\ell}\mid X_{1:\ell-1})}[\log \pi(X_{\ell}\mid X_{1:\ell-1})]
\end{equation}
Our key observation is that a forward pass through an autoregressive model does not just return the probability of one letter, $\pi(X_\ell \mid X_{1:\ell-1})$, it also returns the probability of all the other letters occurring at that position, $\pi(b \mid X_{1:\ell})$ for all $b \neq X_\ell$, at no extra computational cost.
It is also tractable to compute $q_\theta(X_\ell \mid X_{1:\ell-1}) = q_\theta(X_{1:\ell})/q_\theta(X_{1:\ell-1})$. 
Thus, we can compute the inner expectation in \Cref{eqn:ar-rb} analytically rather than by Monte Carlo, Rao-Blackwellizing the estimate.

\subsection{Quantization}

After training, we impose synthesis constraints with a discretization method.
This is analogous to post-training quantization of neural networks: we are converting the model parameters to lower precision so that sampling can be run on different hardware \citep{Nagel2021-ln,Xiao2023-li}.

We discretize each $\theta_\ell^m$ to come from a finite catalog $\mathcal U$ by minimizing a distance $d$: $\tilde\theta_\ell^m = \mathrm{arg\,min}_{\theta\in\mathcal U}d(\theta, \theta^m_\ell)$.
In practice, we found $d(\theta, \theta')=\mathrm{KL}(\theta||\theta')$ to work best.
This effectively "rounds" the parameters to the closest physically achievable value.
On some synthesizers we can expand the catalog by adding $K$ pre-set mixtures, $\psi_k \in \Delta^4$.
To discretize variational synthesis models to run on this hardware, we first design the mixtures to minimize the total distance,
\begin{equation} \label{eqn:distance-obj}
    \psi_\star = \argmin{\psi} \sum_{\ell, m}\underset{\theta\in\mathcal U_\psi}\min d(\theta, \theta_{\ell}^m)
\end{equation}
This objective is piecewise linear, so we apply Adam to optimize. Then, we discretize using $\mathcal U_{\psi_\star}$.

To design DNA that encodes proteins, we train and discretize in amino acid space.
Chemically, most synthesizers use mixtures of nucleotides rather than trinucleotides, so we must translate nucleotide mixture constraints into amino acid mixture constraints. 
Define $\mathcal{T}(\mathcal{U}) \subset \Delta^{21}$ to be the set of $|\mathcal{U}|^3$ distributions over amino acids that can be encoded by nucleotide mixtures in $\mathcal{U}$ \citep{Weinstein2022-sw}.
We discretize by setting $\tilde\theta_\ell^m = \mathrm{arg\,min}_{\theta\in \mathcal{T}(\mathcal U)}d(\theta, \theta^m_\ell)$. 
When pre-set mixtures are available, we optimize $\mathcal{U}_\psi$ by  $\psi_\star = \mathrm{arg\,min}_\psi \sum_{\ell, m} \min_{\theta\in\mathcal T (\mathcal U_\psi)} d(\theta, \theta_{\ell}^m)$

\section{Empirical Results} \label{sec:results}

We study cVS \textit{in silico} and \textit{in vitro}.
We consider four design challenges, creating enzymes, peptides, antibodies and regulatory elements. 
These problems involve both short and long sequences, both proteins and DNA, and both forward and reverse KL objectives.
We measure the performance of the cVS training algorithm, relative to the previous EM \citep["EM VS"][]{Weinstein2022-sw} and reinforcement learning \citep[PGLD][]{Sussex2026-am} algorithms, while holding the synthesis constraints fixed.
We focus mainly on settings where the user only has access to low capacity synthesis hardware, with no pre-set mixes and tens of wells $M$ \citep{Sussex2026-am}, rather the high capacity models with pre-set mixes and thousands of wells demonstrated in \citet{Weinstein2026-vs}.
In this low capacity regime, differences in training algorithm can have especially large effects.
With the reverse KL objective, >99\% of runtime came from the prior and reward models, rather than the variational synthesis model. For the forward KL, models trained in less then 5 minutes on an A100 or H100 GPU. 
Across all design challenges, we find cVS substantially advances the Pareto frontier of quality and diversity. 
Finally, we confirm that cVS's strong \textit{in silico} performance is maintained \textit{in vitro}.

\subsection{Designing full length enzymes}
\begin{figure}[t]
    \centering
    \begin{tabular}{@{}m{0.1\textwidth}m{0.9\textwidth}@{}}
    (a) & \includegraphics[width=\linewidth]{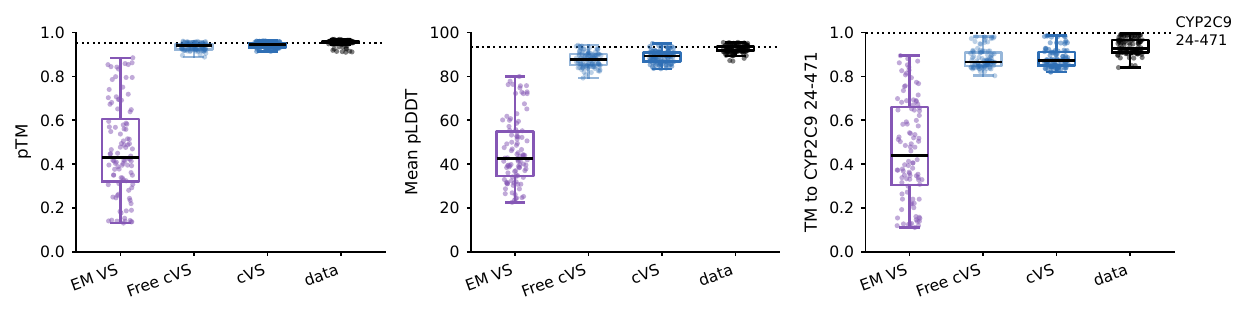} \\[2pt]
    (b) & \includegraphics[width=\linewidth]{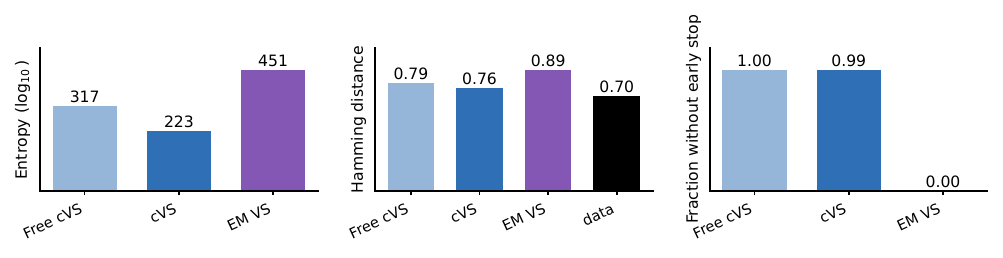} \\[2pt]
    (c) & \includegraphics[width=\linewidth]{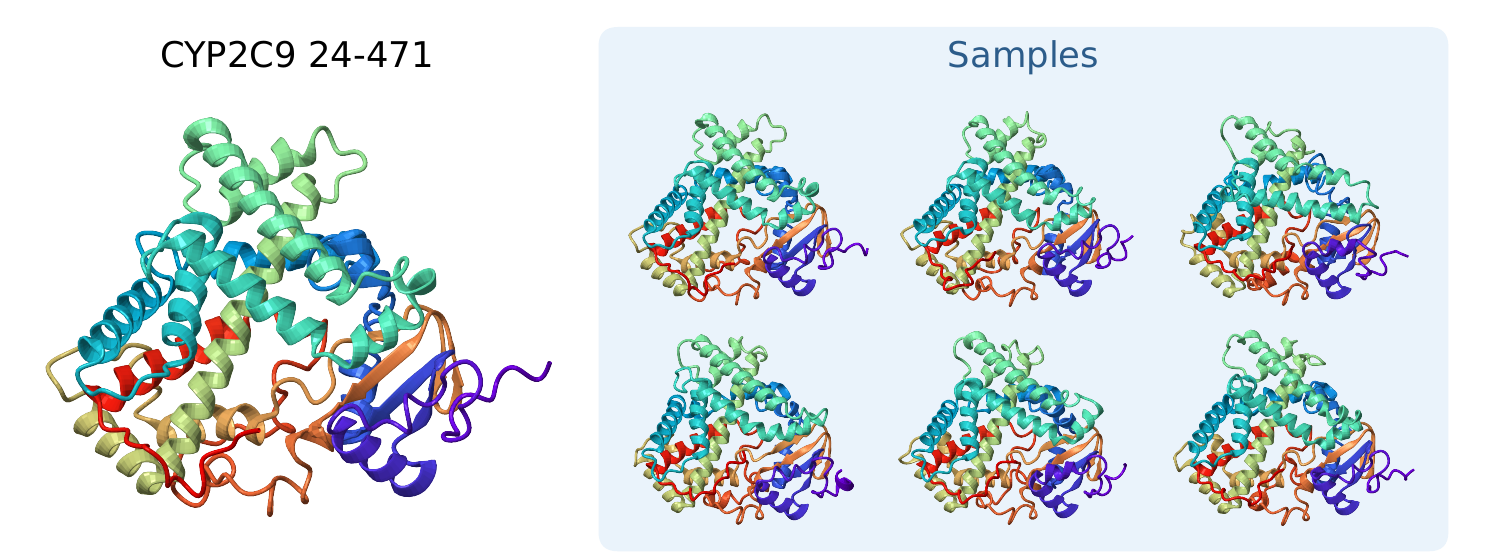} \\
    \end{tabular}
    \caption{\textbf{Enzyme design (Cytochrome P450).} (a) Quality: ESMFold's pTM and pLDDT, and TM-score to the predicted CYP2C9 structure. We compare cVS pre- and post-quantization (free cVS, cVS) to EM VS and data. (b) Diversity: entropy of the library (left), and mean Hamming distance (center); higher is more diverse on both. The EM library had many internal stop codons (right); the other evaluations are on samples with no stops. (c) Predicted structure of independent samples from the cVS library, compared to that of the human cytochrome P450 2C9.}
    \label{fig:enzyme_summary}
\end{figure}

We study cytochrome P450 (CYP), an enzyme involved in drug metabolism that has been reengineered to catalyze diverse reactions \citep{Coelho2013-uj}. 
We sought to design a library covering CYP's evolutionary diversity while maintaining its structure. 
We use the forward KL objective, to cover the evolutionary family.
We design the enzyme's full length, a long-range design problem.

\paragraph{Setup.}
We started from an alignment of the human cytochrome P450 2C9 (CYP2C9) to evolutionarily related sequences \citep{Notin2023-fx}.
We use this alignment as training data, treating gaps as missing data, and ignoring insertions relative to the human sequence.
The alignment has length 434 amino acids.
We consider low capacity synthesis with $M=16$ wells and $\mathcal{U}$ the equal nucleotide mixtures.
We optimize the forward KL (\Cref{eqn:fwd-kl}).
Details in \Cref{apx:enzyme}.

\paragraph{Results.}
We compare cVS to EM VS, which uses the same forward KL objective \citep{Weinstein2022-sw}.
The EM designs are poor: because of the long sequence length and strong synthesis constraints, there are internal stop codons in almost every sample, and only $2 \times 10^{-6}$ do not have one.
We examined the structure of the cVS designs, predicted by ESMFold \citep{lin2023evolutionary}.
Samples from the cVS library produce high confidence structures: the pTM and pLDDT are nearly as high as evolutionary sequences (\Cref{fig:enzyme_summary}a). They are structurally similar to the human enzyme, quantitatively (TM above 0.8) and qualitatively (\Cref{fig:enzyme_summary}c).
The designs are diverse, with similar mean Hamming distance to the natural sequences (\Cref{fig:enzyme_summary}b).
Overall, cVS enables long range stochastic synthesis design, even in the presence of strong constraints.

\begin{figure}
    \centering
    \includegraphics[width=0.8\linewidth]{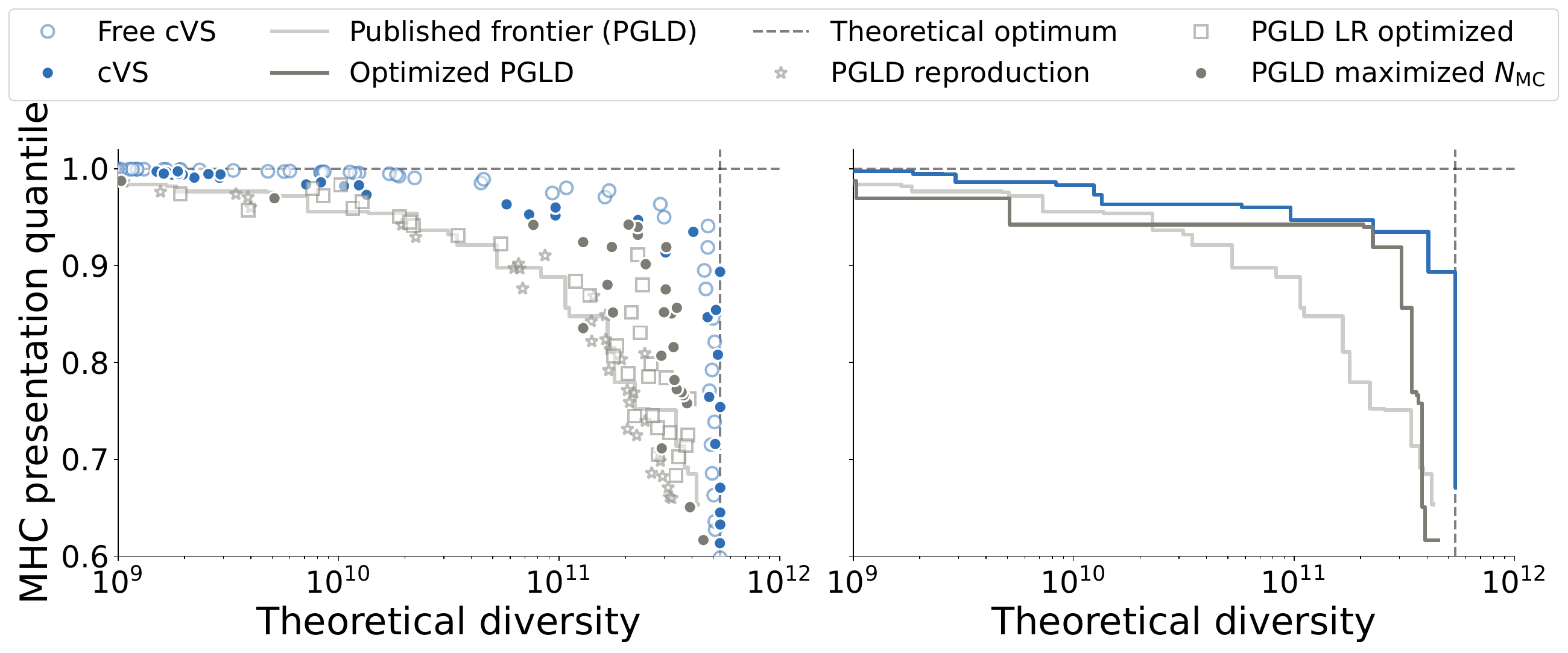}
    \caption{\textbf{Quality-diversity Pareto frontier on benchmark peptide design.}  Y-axis: mean binding quantile of peptides in the designed library, compared to a reference distribution. X-axis: "theoretical diversity". We compare cVS (blue) to PGLD (gray). We show the raw data from \citep{Sussex2026-am} (published frontier) and our reproduction (stars), with learning rate (LR) optimized and maximized $N_{MC}$. We compare to cVS designs before (free cVS) and after quantization (cVS). Left: Reproduction and hyperparameter improvement of PGLD compared to cVS. Right: Pareto frontiers for PGLD (published, optimized), and for cVS (post-quantization).}
    \label{fig:benchmark-summary}
\end{figure}

\subsection{Designing \hla-presented peptides}  
Human leukocyte antigen (HLA) molecules present short peptides on the cell surface, where they can be recognized by the adaptive immune system. 
We study a benchmark from \citet{Sussex2026-am} where the goal is to design peptides presented by \hla. They consider a reverse KL objective, with the reward specified by a binding predictor, and access only to low capacity synthesis hardware.

\paragraph{Setup.}
Public code for PGLD is unavailable, so we reproduce the method and evaluation.
The reward is a neural network trained to predict peptide binding from amino acid sequence, MHCflurry 2.0 \citep{O-Donnell2020-mb}. 
The prior is uniform over length 9 amino acid sequences, with no stop codons except possibly in the last position.
PGLD assumes equal nucleotide mixtures $\mathcal{U}$ and uses $M=32$ wells for this benchmark, i.e. low capacity synthesis; we use cVS with the same constraints. The evaluation is the Pareto frontier between expected reward and "theoretical diversity", defined as the number of unique sequences in the library when amino acids with probabilities less than 3\% at each position are removed.
To evaluate cVS's Pareto frontier, we sweep a hyperparameter $\alpha$ specifying the balance of the reward and the prior, $\alpha\, \Eb_{q_\theta}[r(X)] - \kl(q_\theta \| \pi)$.
Details in \Cref{apx:epitopes}.
Note we do not use "theoretical diversity" in later evaluations: its choice in \citet{Sussex2026-am} is motivated by limitations on downstream assays and analysis that require testing multiple copies of the same sequence, but these limitations are unnecessary given recent advances in learning from variational synthesis \citep[LeaVS, LIFT][]{Weinstein2026-eg,Weinstein2025-fv}.

\paragraph{Reproduction.} Our reproduction of PGLD closely matched the published Pareto frontier (\Cref{fig:benchmark-summary}).
Then, we applied the same hyperparameter optimization method as for cVS, sweeping the learning rate. PGLD has an additional hyperparameter, $N_{MC}$, which we observed should be set to larger values than originally proposed. These changes improved PGLD (\Cref{fig:benchmark-summary}).

\paragraph{Results.} We trained the same synthesis model with cVS, which substantially advanced the Pareto frontier, especially in the high diversity regime (\Cref{fig:benchmark-summary}). 
Post-training discretization imposed a relatively modest cost on performance: there is not a large gap between free cVS and the quantized cVS.
Performance is robust to hyperparameters of the training algorithm, with cVS showing smaller sensitivity than PGLD (\Cref{fig:benchmark_eval_scan,fig:benchmark_hypers}).
Overall, cVS achieves state-of-the-art performance training variational synthesis models with a reverse KL objective.

\subsection{Designing TCR mimicking antibodies}
We evaluated cVS on lab-in-the-loop antibody design \citep{Frey2025-cq}. 
We aim to design TCR mimicking (TCRm) antibodies that specifically bind peptides presented on an HLA molecule (pHLAs) \citep{Klebanoff2023-bx}. 
pHLAs are exceptionally challenging targets, since the peptides are in a dynamic complex with the HLA, but they are highly sought for therapeutics, since they allow immunotherapies to be directed against intracellular antigens \citep{Hsiue2021-ci,salzler2025car,Yarmarkovich2021-kw,Liu2025-hp,Householder2025-js}.

\paragraph{Setup.}
We start with a transformer protein-protein interaction model, trained on data collected by screening an initial variational synthesis scFv library against a panel of 100 pHLA targets in a large scale human display system \citep{Weinstein2026-vs,Weinstein2026-eg,Unknown2026-og}.
We use cVS to design the next round library for testing.
The reward $r(x)$ is specified by the scFv-pHLA interaction model; it is the maximum expected binding counts across target pHLAs.
The prior $\pi(x)$ is an autoregressive transformer trained on the initial library.
We start by assuming low capacity synthesis, using $M=32$ wells and equal nucleotide mixtures.
We design the CDRH3 region, with a fixed length of $L=15$ amino acids. 

We first pre-train with a forward KL objective, using samples from the prior reweighted by the reward to approximate the target distribution, $\tilde p(x) \propto \pi(x) \exp(r(x))$. 
Then, we fine-tune with the reverse KL.
To fairly evaluate PGLD we use the same pre-training and fine-tuning; note forward KL pre-training was not originally proposed for PGLD but substantially improved its performance (\Cref{fig:mesa_pretrain}).
EM VS only allows forward KL pre-training.
We evaluate designs' quality, measured by the mean reward across the library $\Eb_{q_\theta}[r(X)]$, and diversity, measured by the KL divergence to the prior, $\kl(q_\theta \| \pi)$.
We sweep the reward weight $\alpha$ to explore the Pareto frontier.

\paragraph{Results.}

\begin{figure}
    \centering
    \begin{subfigure}{0.32\textwidth}
    \includegraphics[width=\linewidth]{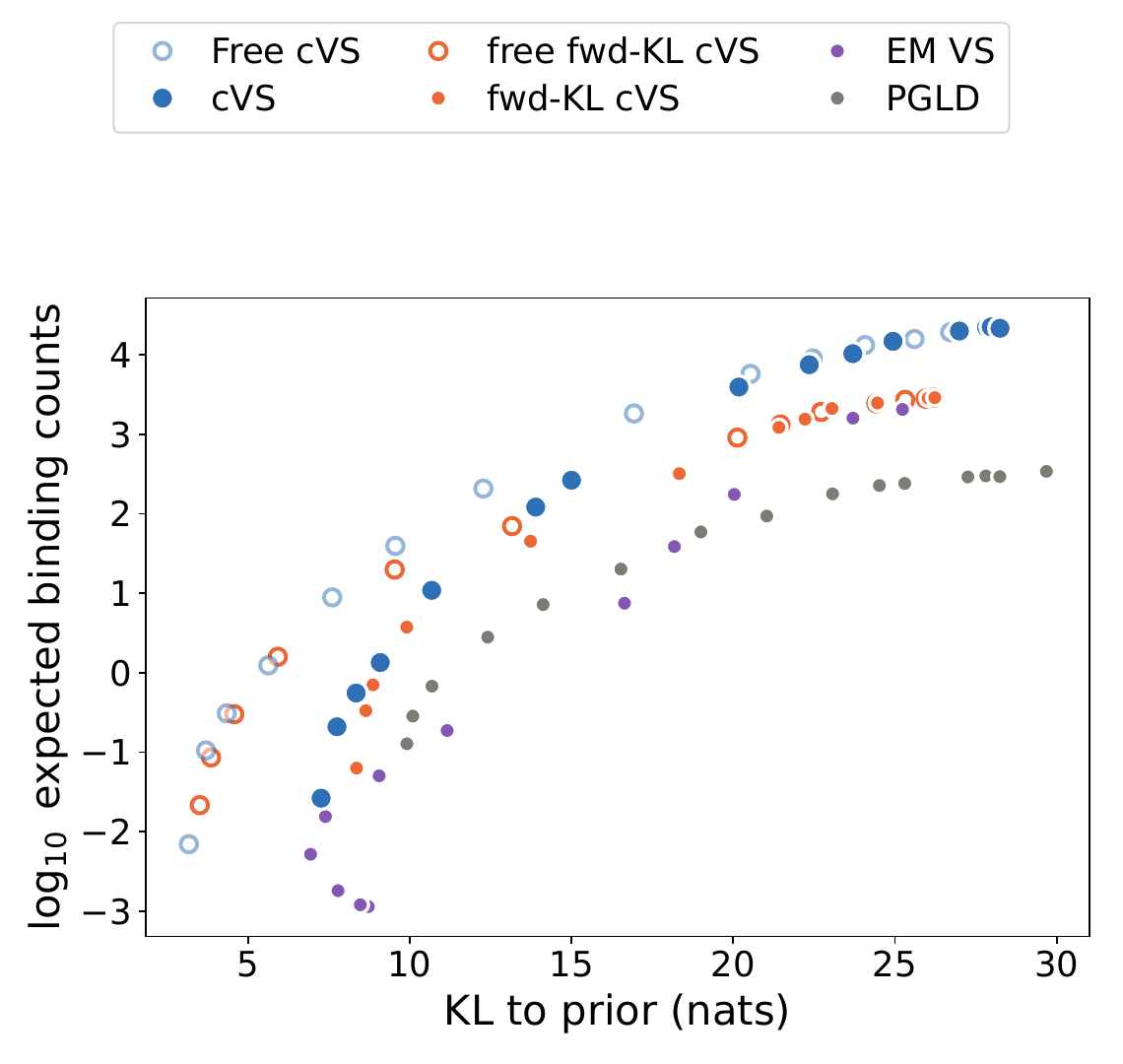}
    \caption{}\label{fig:mesa_summary}
    \end{subfigure}
    \begin{subfigure}{0.32\textwidth}
    \includegraphics[width=\linewidth]{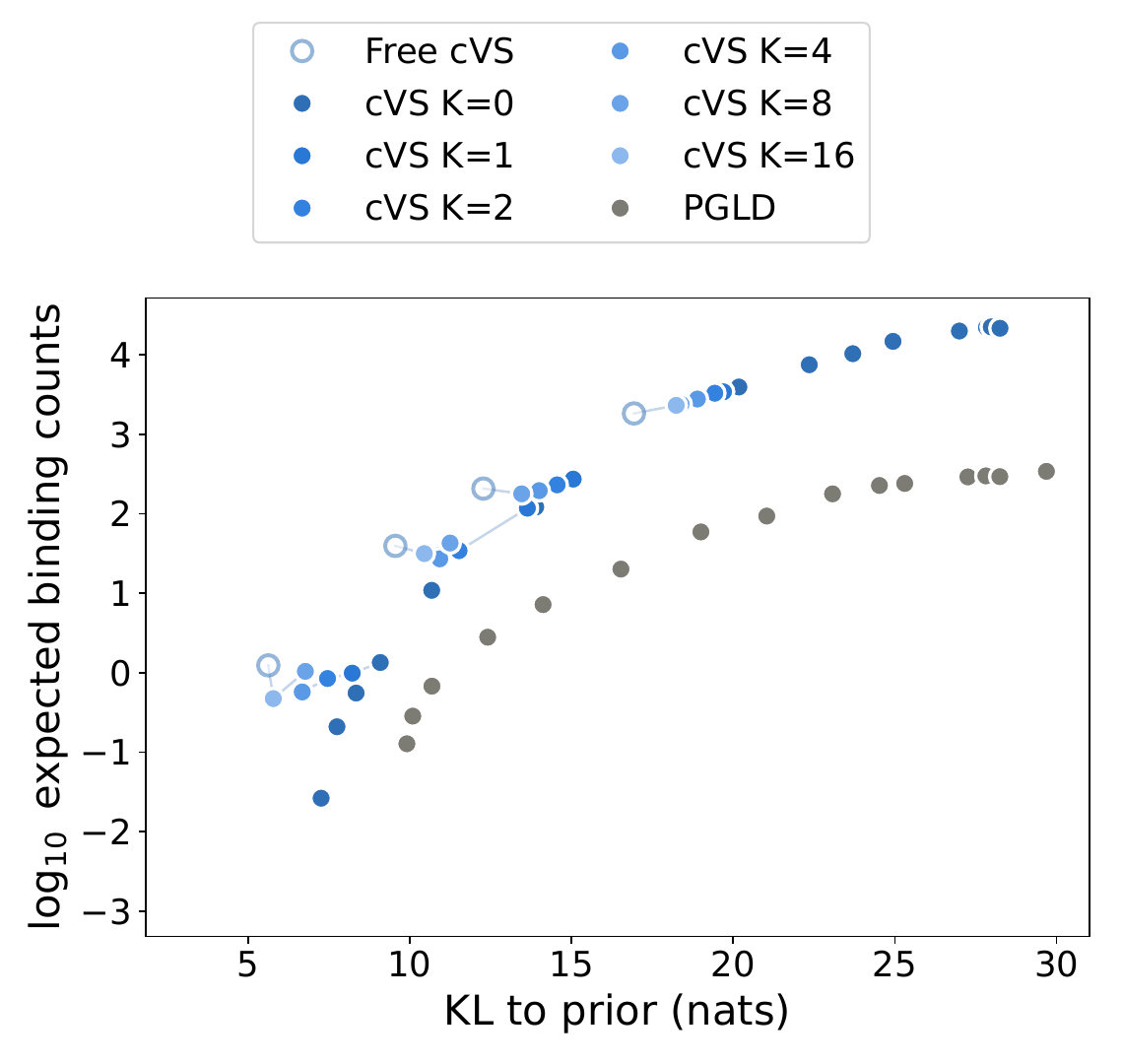}
    \caption{}\label{fig:mesa_mix_scan}
    \end{subfigure}
    \begin{subfigure}{0.32\textwidth}
    \includegraphics[width=\linewidth]{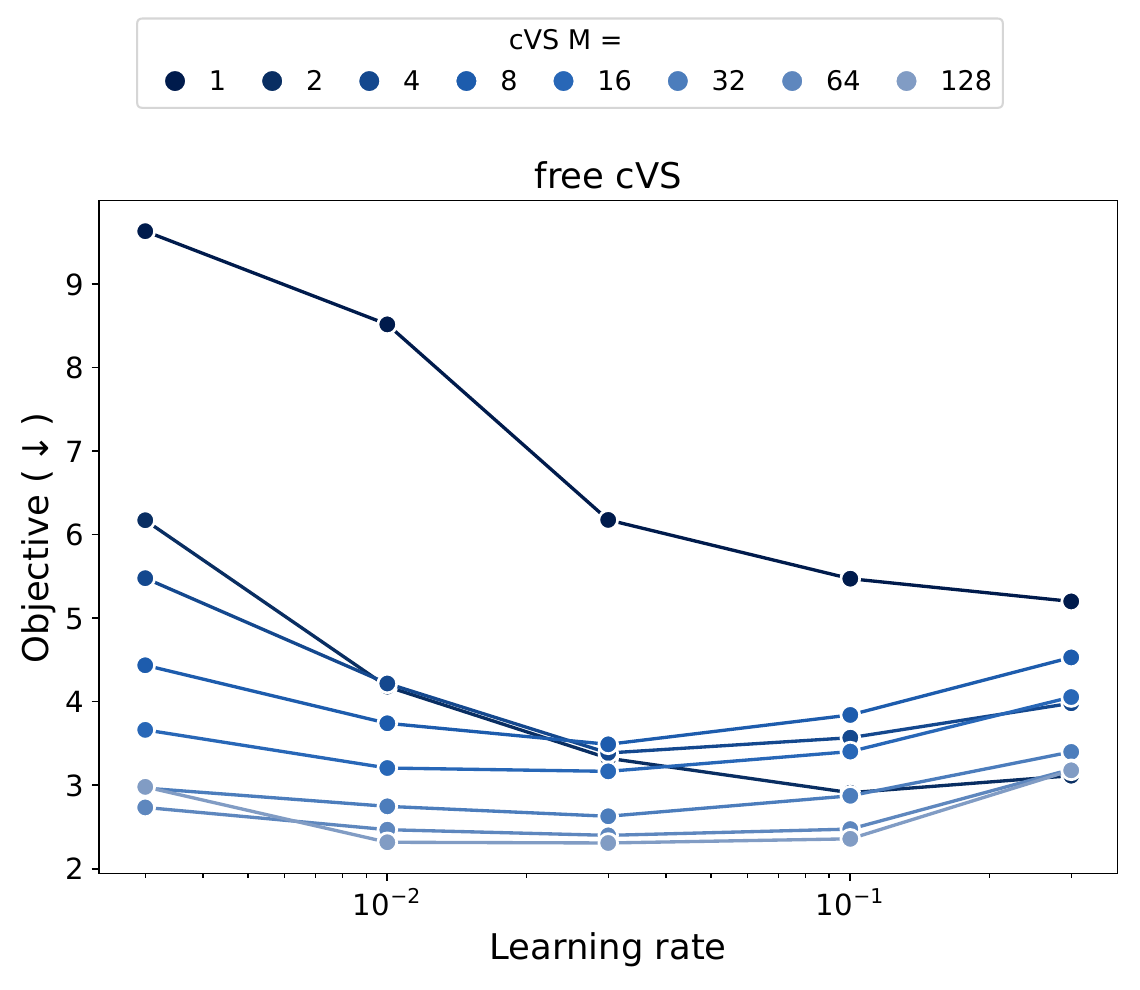}
    \caption{}\label{fig:mesa_M_panel}
    \end{subfigure}
    \caption{\textbf{scFv CDRH3 library design (TCR mimics).} (a) Pareto frontier of quality (mean predicted binding counts against a pHLA target) versus diversity (KL to human repertoire prior). 
    (b) Performance with changing synthesis hardware and wetware, increasing the number of pre-set mixes available in $\mathcal{U}_{\psi_{1:K}}$.
    (c) Performance of free cVS with increasing wells $M$. }
    \label{fig:mesa_fwd}
\end{figure}

cVS Pareto dominates both EM VS and PGLD, achieving higher reward and higher diversity (\Cref{fig:mesa_summary}).
cVS's performance is robust to hyperparameters of the training algorithm including the Adam hyperparameters, and is less sensitive than PGLD (\Cref{fig:mesa_hypers,fig:mesa_eval_scan}).

We evaluated the pre-training. 
Optimizing the forward KL, cVS outperformed EM VS (\Cref{fig:mesa_summary}). 
Removing the pre-training, and only using reverse KL, harmed performance (\Cref{fig:mesa_pretrain}).
Removing the fine-tuning also harmed performance (cVS vs fwd-KL cVS, \Cref{fig:mesa_summary}).
This held even when the total reward evaluations was held fixed (\Cref{fig:mesa_matched_eval}).
In sum, pre-training followed by fine-tuning produces the best performance.

We ablated our gradient variance reduction (\Cref{fig:mesa_rao_blackwell}). We see only minor performance drop, suggesting cVS's main advantage comes from the continuous relaxation plus quantization.

We investigated alternative diversity metrics. 
We retrained using a uniform prior, and evaluated the Shannon entropy.
cVS Pareto dominates the reward vs. entropy frontier (\Cref{fig:mesa_shannon_entropy}).
We then estimated designs' kernelized 2-Renyi entropy, which measures diversity at different scales, corresponding to different kernel bandwidths \citep{Sanchez-Giraldo2012-qd}.
Both cVS and PGLD find modes that are spread out in sequence space, but cVS's modes are wider (\Cref{fig:mesa_shannon_entropy}).

We next explored higher capacity synthesis, using additional nucleotide mixtures.
We expanded the catalog $\mathcal{U}$ with pre-set mixes $\psi_{1:K}$, optimized the mixture choice (\Cref{eqn:distance-obj}), and re-quantized the free cVS model.
Performance improves systematically, more closely approximating the free cVS value (\Cref{fig:mesa_mix_scan}).
(Note the infinite $K$ limit does not necessarily approach free cVS asymptotically, since not all amino acid distributions can be made by drawing each nucleotide in the codon independently.)

Next, we expand synthesis capacity by increasing the number of reaction wells $M$, i.e. the number of components in the mixture model.
Performance of free cVS improves smoothly (\Cref{fig:mesa_M_panel}), and translates into better performance post-discretization (\Cref{fig:mesa_M_scan}).
Heuristically, we expect a stable McKean-Vlasov-style interacting particle limit for free cVS, since it performs gradient based optimization of the reverse KL with a mixture model, albeit one that is non-Gaussian \citep{Lambert2022-fo,Wild2023-yv}. 
Indeed, as $M$ increases, the optimal learning rate decreases and performance approaches a limiting value (\Cref{fig:mesa_M_scan}).

Overall, cVS offers state-of-the-art performance on a lab-in-the-loop biologics design problem.

\subsection{Engineering regulatory elements}

\begin{figure}
    \centering
    \includegraphics[width=\linewidth]{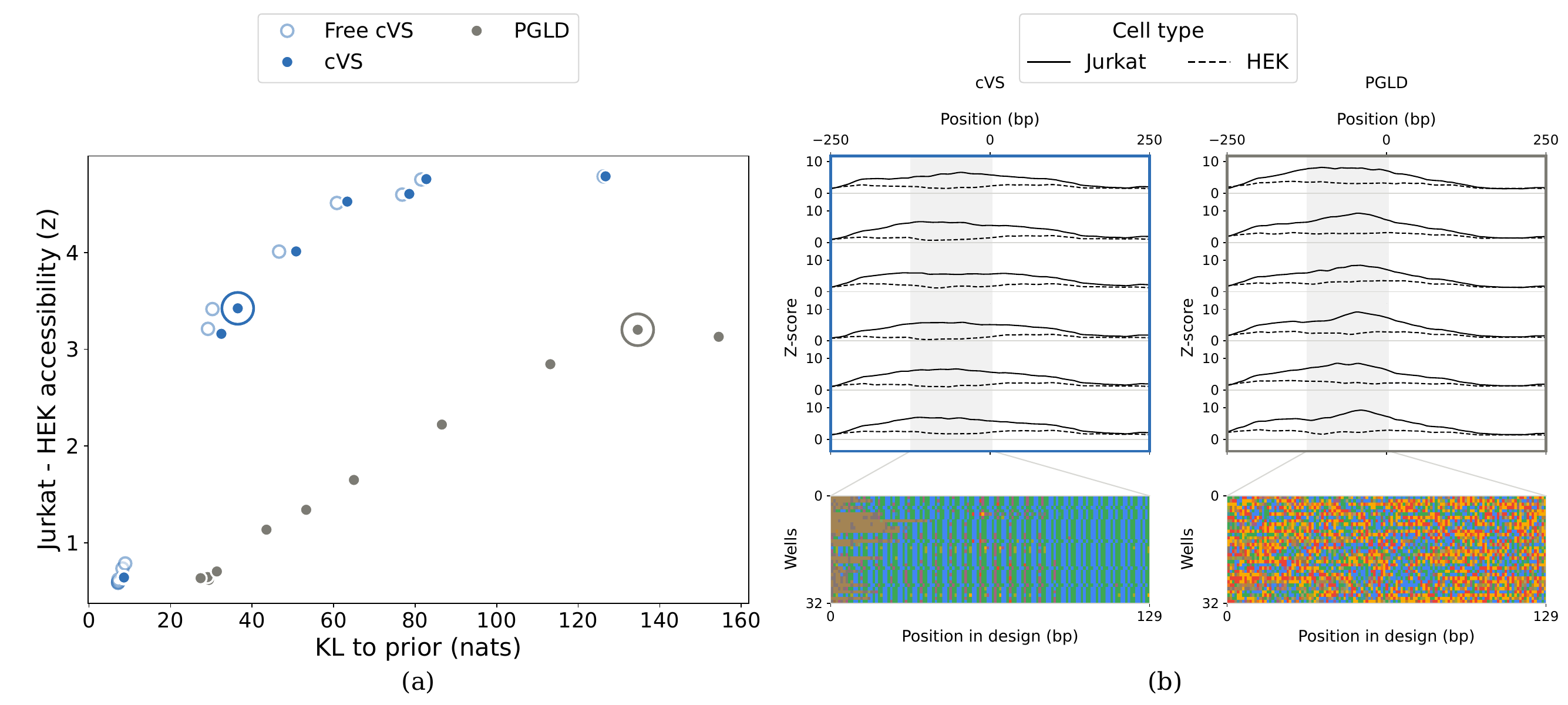}
    \caption{\textbf{Regulatory DNA design (EF1$\alpha$).} 
    (a) Quality-diversity Pareto frontier, evaluating the average difference in cell-type accessibility versus the KL to the human genome prior. (b) Predicted accessibility of sampled sequences (above) from the learned synthesis models (below). Positions in each well are colored by the nucleotide mixture.}
    \label{fig:atac_summary}
\end{figure}

We next apply cVS to regulatory DNA. We aim to reengineer EF1$\alpha$, a human promoter that drives strong expression across diverse cell types. EF1$\alpha$ is widely used in CAR-T cell therapy, but it risks expression in off-target cell types \citep{Nyberg2026-ob}.
We sought to decrease EF1$\alpha$'s activity outside of T cells while maintaining high expression within T cells. 

\paragraph{Setup.}
We first established a reward model. We train a BPNet-style convolutional neural network, with context size 2048, to predict ATAC-seq data \citep{Avsec2021-xm}. We use data collected from Jurkat, a T cell line, and HEK cells, as an off-target cell line \citep{Zou2024-eq}.
The reward $r(x)$ is the predicted difference in mean chromatin accessibility across a 300 base pair window.
For the prior $\pi(x)$ we use a generative model of genome sequences. 
We fine-tune MarinDNA, an autoregressive evolutionary genomic language model, on the same human genome regions used to train the reward model \citep{Benegas2026-ae}.
The variational synthesis model uses $M=32$ wells and the equal nucleotide constraint. We design a 150 bp region near the start of EF1$\alpha$.
We pre-train via the forward KL and fine-tune via the reverse KL, for both cVS and PGLD.
Details in \Cref{apx:promoter}.

\paragraph{\textit{In silico} results.}
cVS provides a substantial advance in the Pareto frontier, designing libraries with high cell-type specific accessibility that stay close to the human genome prior (\Cref{fig:atac_summary}a).
In comparison to the protein design problems, the penalty from quantization is small, with only a small gap between free cVS and quantization to equal nucleotide mixtures.

We visualized the synthesis model $q_{\theta_\star}(x)$, observing that cVS learns structured sequence designs with tiled motifs, in contrast to PGLD (\Cref{fig:atac_summary}b).
Reengineering EF1$\alpha$ activity is particularly challenging: when we retrained in a random genomic context, we again observed a substantial improvement from cVS, but also greater variety in cVS's designs (\Cref{fig:atac_summary_random}).

\begin{figure}
    \centering
    \begin{subfigure}{0.22\textwidth}
    \centering
    \includegraphics[width=0.9\linewidth]{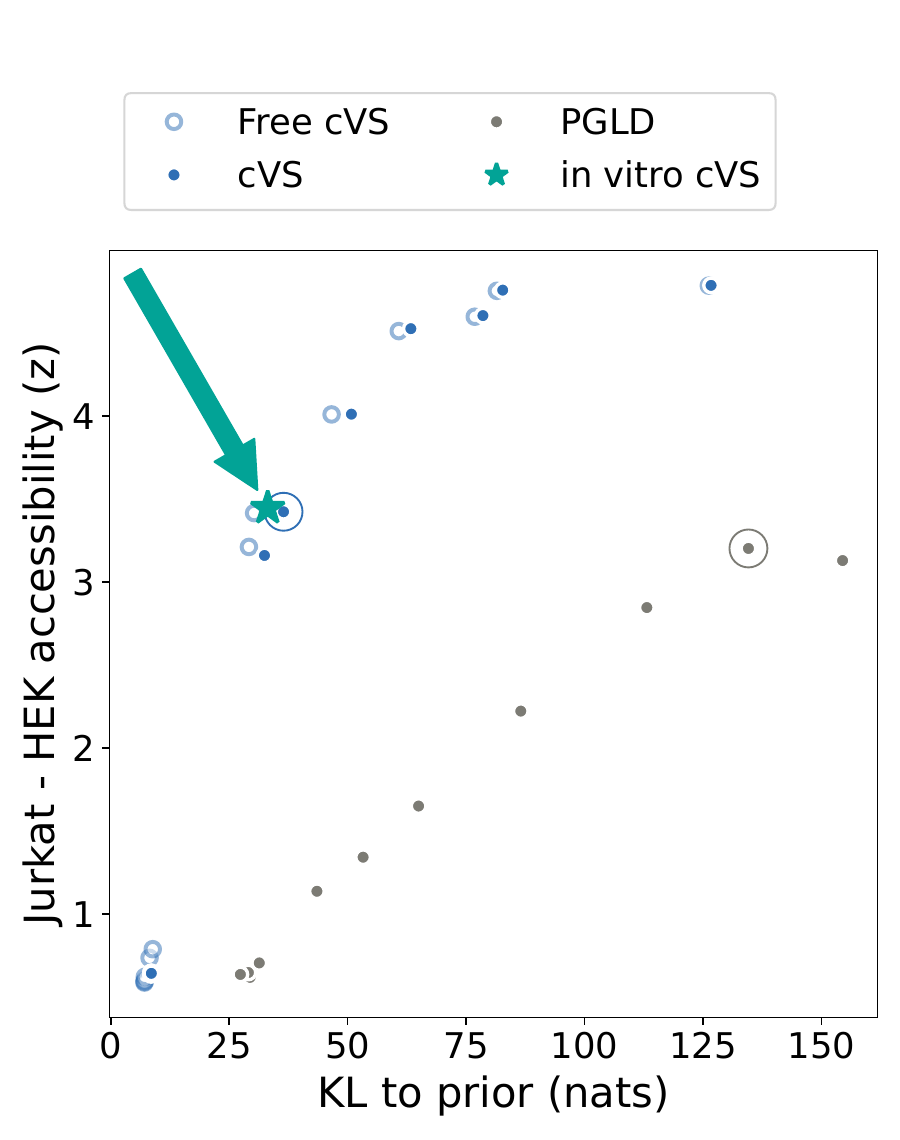}
    \caption{Pareto frontier.}\label{fig:atac_in_vitro_frontier}
    \end{subfigure}
    \begin{subfigure}{0.22\textwidth}
    \includegraphics[width=\linewidth]{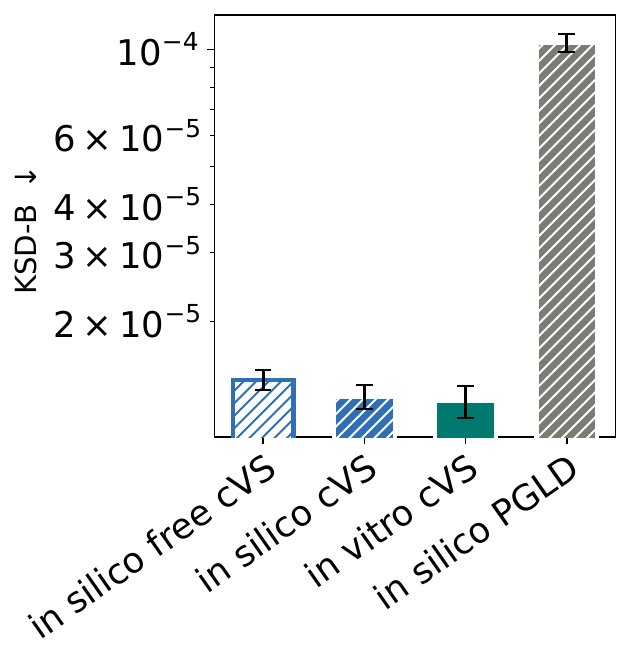}
    \caption{KSD-B}\label{fig:atac_in_vitro_ksdb}
    \end{subfigure}
    \begin{subfigure}{0.22\textwidth}
    \includegraphics[width=\linewidth]{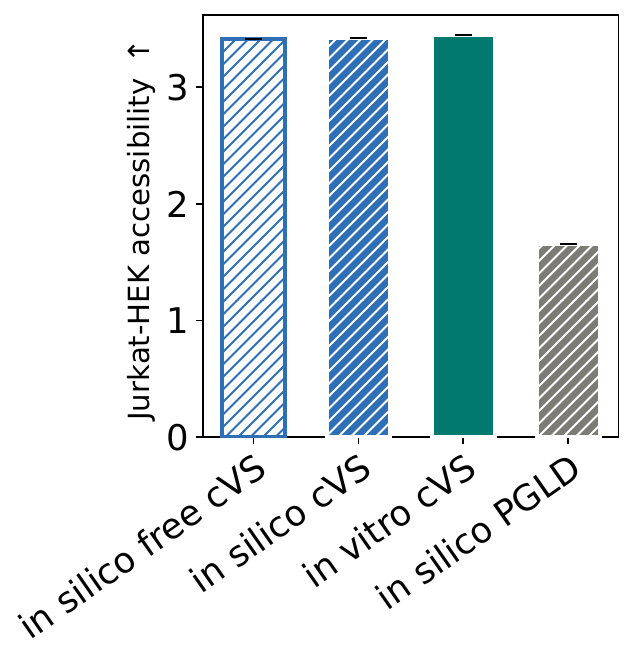}
    \caption{Average reward}\label{fig:atac_in_vitro_reward}
    \end{subfigure}
    \begin{subfigure}{0.22\textwidth}
    \includegraphics[width=\linewidth]{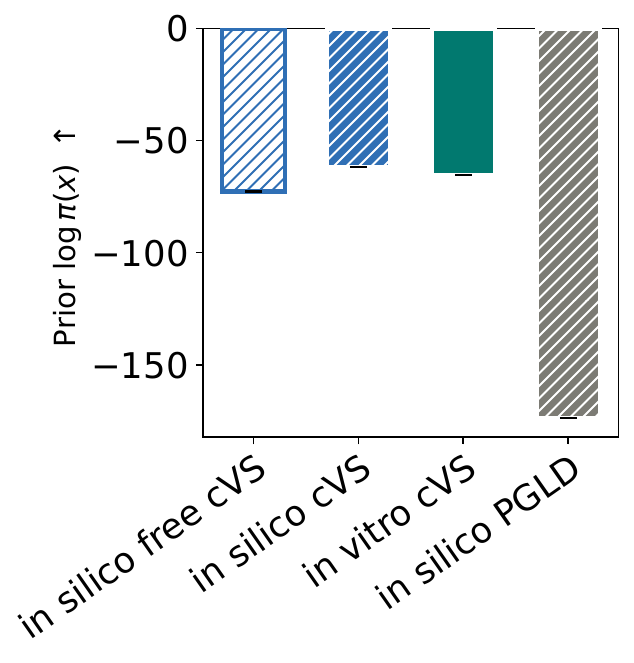}
    \caption{Average (log) prior}\label{fig:atac_in_vitro_prior}
    \end{subfigure}
    \begin{subfigure}{0.2\textwidth}
    \includegraphics[width=\linewidth]{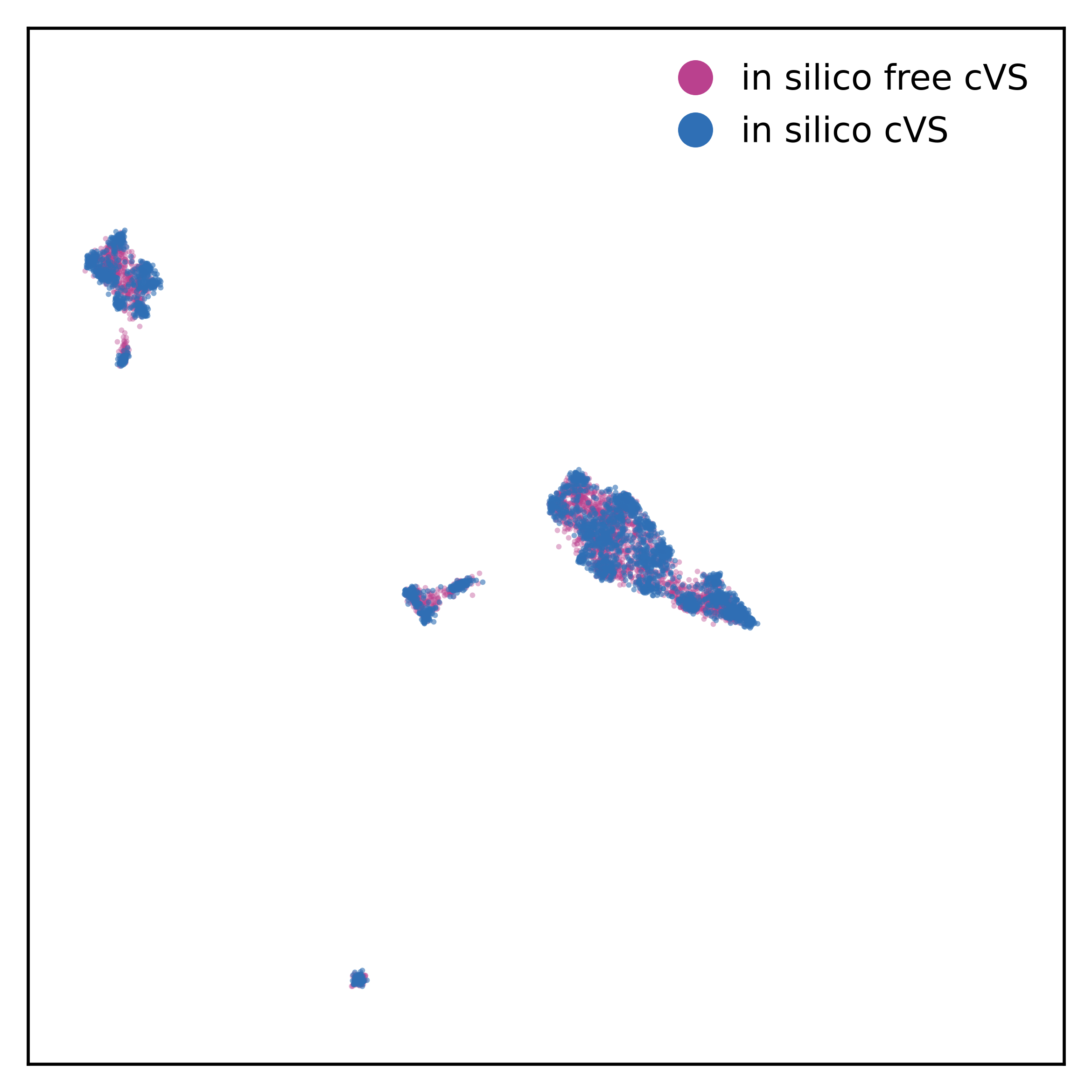}
    \caption{}\label{fig:atac_in_vitro_umap_free_insilico}
    \end{subfigure}
    \begin{subfigure}{0.2\textwidth}
    \includegraphics[width=\linewidth]{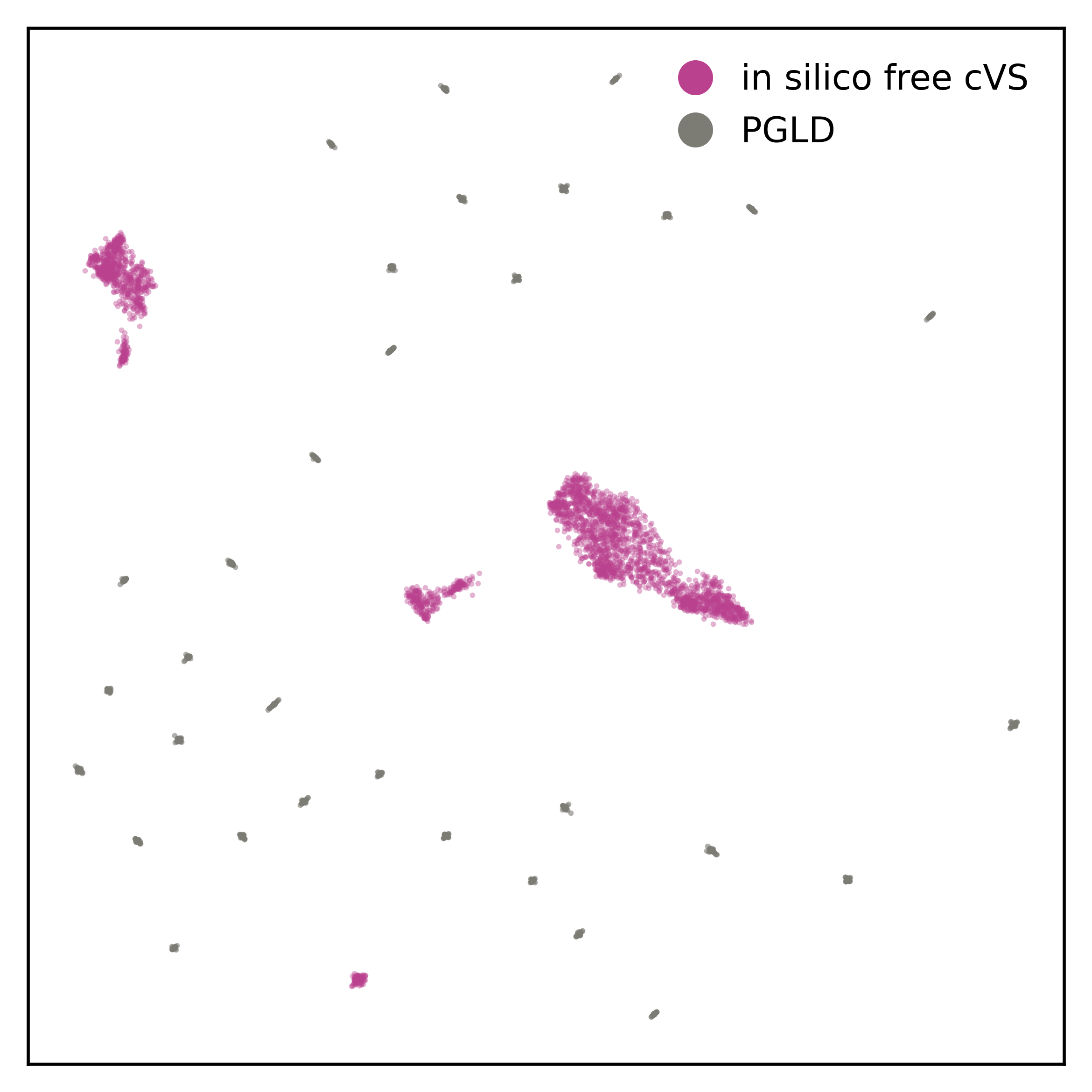}
    \caption{}\label{fig:atac_in_vitro_umap_free_pgld}
    \end{subfigure}
    \begin{subfigure}{0.2\textwidth}
    \includegraphics[width=\linewidth]{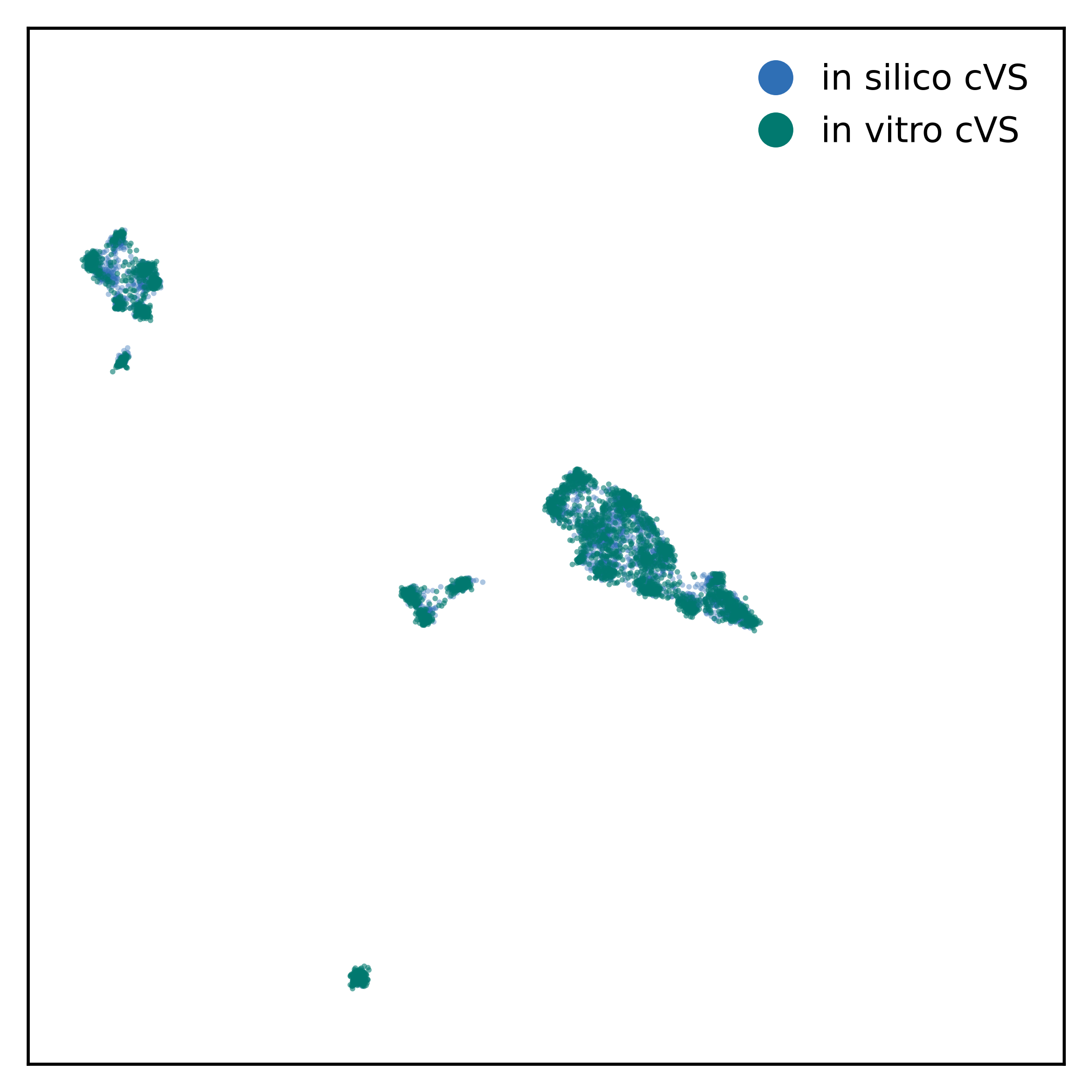}
    \caption{}\label{fig:atac_in_vitro_umap_free_invitro}
    \end{subfigure}
    \caption{\textbf{\textit{In vitro} validation of the EF1$\alpha$ promoter library}: (a) Estimated position of the \textit{in vitro} library on the Pareto frontier.
    (b) KSD-B goodness-of-fit to the reverse-KL target $\pi(x)\exp(\alpha r(x))$. Mean and standard error (SEM) from five independent redraws of 1000 samples from the \textit{in silico} model or the \textit{in vitro} sequencing data. (c) Average reward and SEM.
    (d) Average prior log likelihood and SEM.
    (e-g) Low-dimensional representation of the samples from the prior model (pink) overlaid with the samples from each of the evaluated models. 
    }
    \label{fig:atac_in_vitro}
\end{figure}
\paragraph{\textit{In vitro} results.} We synthesized samples from a trained variational synthesis model $q_{\theta_\star}(x)$, by running the learned stochastic synthesis protocol in the lab.
We achieved a yield of 1.6 nM, or $9.6 \times 10^{14}$ molecules.
To evaluate library quality, we sequenced a random subset, obtaining 28 million assembled paired-end reads.
We then checked how well these samples matched the target distribution, the prior tilted by the weighted reward, $\tilde p(x) \propto \pi(x) \exp(\alpha r(x))$.
Evaluating synthesis is nontrivial: we only have samples from the \textit{in vitro} distribution $q_{\textit{iv}}(x)$, not likelihoods, and we only know the target distribution $\tilde p$ up to a normalizing constant, without access to samples. 
We can estimate the position of the \textit{in vitro} library on the Pareto frontier, but only by an upper bound on the entropy, $-\mathbb{E}_{q_{\textit{iv}}(x)}[\log q_{\theta_\star}(x)] \ge -\mathbb{E}_{q_{\textit{iv}}(x)}[\log q_{\textit{iv}}(x)]$, which provides an optimistic estimate of diversity (\Cref{fig:atac_in_vitro_frontier}).
We therefore use the KSD-B, an extension of the kernelized Stein discrepancy to discrete sequences, which allows comparison of samples to an unnormalized distribution \citep{Amin2023-dc,Liu2016-bp,Gorham2017-sd}.
We use a Hamming IMQ kernel, which ensures the divergence can detect arbitrary nonparametric mismatch \citep{Amin2023-dc,Amin2025-cg}.

We find the \textit{in vitro} cVS designs maintain the same quality as the \textit{in silico} cVS designs, achieving a small KSD-B (\Cref{fig:atac_in_vitro_ksdb}).
Its advantage is robust to hyperparameters of the KSD-B kernel (\Cref{fig:atac_in_vitro_ksd_sensitivity}).
\textit{In vitro} cVS also maintains high reward (\Cref{fig:atac_in_vitro_reward}) and prior likelihood (\Cref{fig:atac_in_vitro_prior}). Qualitatively, we can see the effects of quantization, but still observe a close match between \textit{in vitro} cVS and \textit{in silico} cVS and free cVS (\Cref{fig:atac_in_vitro}d-g).
Across all metrics, the \textit{in vitro} cVS library substantially outperforms even \textit{in silico} PGLD.
In sum, we synthesize nearly a quadrillion samples from a generative model with high reward and diversity.

\section{Discussion}

We introduced continuous variational synthesis (cVS), a new approach to training variational synthesis models.
It relies on a continuous relaxation of chemical synthesis constraints, which enables gradient-based training, followed by post-training quantization, which enables synthesis on diverse hardware and wetware.
cVS shows state-of-the-art performance \textit{in silico} on both forward KL and reverse KL objectives, enabling pre-training and fine-tuning variational synthesis models with diverse reward models. It designs both DNA and proteins. 
Its \textit{in silico} performance is maintained \textit{in vitro}.

\paragraph{Limitations.} Synthesizing designs beyond single oligos, such as full length CYP, requires assembly or enzymatic synthesis, which can introduce additional constraints and errors.
The impact of errors depends on the reward model: although we saw minimal impact on chromatin accessibility, other reward models may be more sensitive.

\paragraph{Outlook.} 
cVS allows variational synthesis models to be trained, fine-tuned and deployed like language models. It thus allows ideas and methods originally developed for language modeling to be redeployed to explore biological sequence space, not just \textit{in silico}, but through massive, programmable wet lab experimentation.

\section*{Author contributions}
ANA developed the method and performed all in silico experiments. MGG contributed code for EM VS and synthesis model constraints. MGG planned and implemented the in vitro experiments. AS and MGG analyzed the in vitro experiments. EBW and ENW oversaw the project and advised at all stages. ENW, ANA and AS wrote the paper with input from all authors.

\bibliography{references}
\bibliographystyle{iclr2027_conference}

\appendix

\section{Additional training details} \label{apx:training-techniques}

We implement several other strategies improve training.

\begin{figure}
    \centering
    \begin{subfigure}{0.99\textwidth}
    \includegraphics[width=\linewidth]{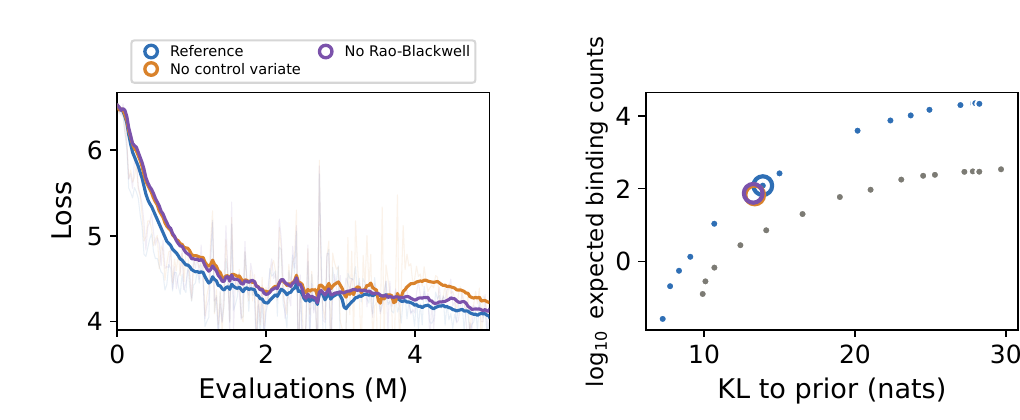}
    \end{subfigure}
    \caption{\textbf{Ablating gradient variance reduction strategies, on the scFv CDRH3 designs.} Removing the prior Rao-Blackwellization or the REINFORCE control variate reduces variance (left) but leads to only a minor change in final performance (right).
    }
    \label{fig:mesa_rao_blackwell}
\end{figure}

\paragraph{Control variates.} Applying tools for learning from human feedback, we use REINFORCE with a customized leave-one-out control variate to reduce the gradient variance \citep{Kool2019-hl,Ahmadian2024-mf}. We ablate the control variate in \Cref{fig:mesa_rao_blackwell}, finding a minor drop in performance.

\begin{figure}
    \centering
    \begin{subfigure}{0.99\textwidth}
    \includegraphics[width=\linewidth]{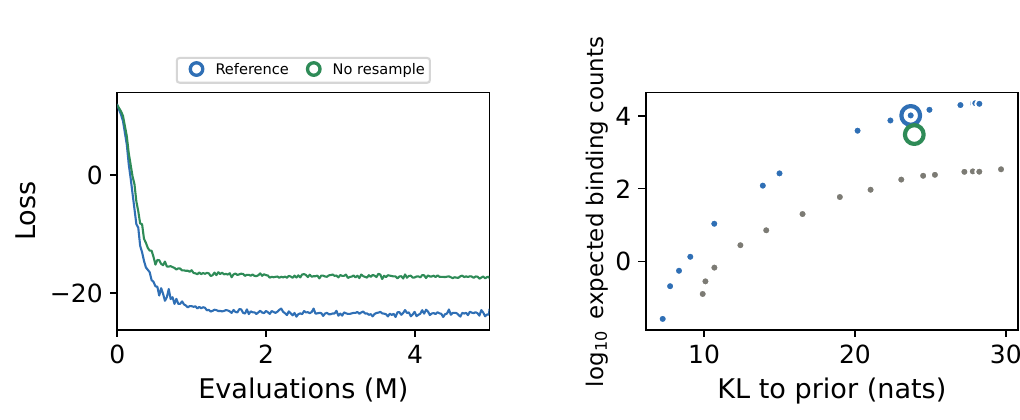}
    \end{subfigure}
    \caption{\textbf{Ablating split-merge resampling of wells, on the scFv CDRH3 designs.} Removing the split-merge step (\Cref{apx:training-techniques}) decreases performance. 
    }
    \label{fig:mesa_resample}
\end{figure}

\paragraph{Split-merge.} We observe that individual wells can get stuck in local maxima. 
We add an auxiliary well weight parameter $w_m$ that describes the fraction of sequences that should be drawn from well $m$, even when the relative concentration is fixed at $1/M$ in practice. We monitor the weights $w_m$ during training, and if they become very low, we resample $\theta_m$ by adding jitter to the parameters $\theta_{m'}$ of the well $m'$ with the highest weight $w_{m'}$.
We ablate this split-merge method in \Cref{fig:mesa_resample}, finding a drop in performance.

\section{Results details}
\subsection{Details on enzyme design}\label{apx:enzyme}

We downloaded the alignment from \citet{Notin2023-fx}.
We ignore insertions relative to the human sequence, and treat gaps as missing data.
Positions 23:470 (0 indexed) were included in the alignment.
The 13 residues 46, 136, 137, 212, 213, 214, 216, 273, 274, 275, 276, 364, 463 were labelled as insertions in the wild type and not included in the alignment. 
We build a generative model of all aligned residues and then put in those 13 insertions (constant positions) back into our designs, i.e. we’re designing the core positions 23:470.
We held out 10\% of alignment sequences for early stopping.
We trained five forward KL cVS models with different learning rates and chose the one with the best held-out loss.

\subsection{Details on \hla epitope repertoire results} \label{apx:epitopes}

In the original PGLD paper they used a fixed learning rate; however the optimal leanring rate can differ across methods and $\alpha$.
To ensure a fair comparison, we swept 5 learning rates for each $\alpha$ and picked the one that achieved the best objective, for both cVS and PGLD.
PGLD also had a hyper-parameter $N_{MC}$ which was set to different values throughout their paper; however, from a variance-reduction perspective, $N_{MC}$ should be set as high as possible.
\Cref{fig:benchmark-summary} shows that sweeping learning rate and maximizing $N_{MC}$ improves the frontier for PGLD.
All experiments are performed with learning rate sweeps and maximal $N_{MC}$.

We optimize the reward $r(x) = \log_{10}(\verb|presentation_percentile|(x)/100)$ where $\verb|presentation_percentile|(x)$ is MHCflurry 2.0's prediction of binding strength relative to a reference distribution. This parameterization sharply penalizes low scores.
The final performance is measured as $1 - \Eb_{q_\theta}[\verb|presentation_percentile|(x)]/100$.
The theoretical diversity is approximated with the Karp-Luby algorithm \citep{Karp1989-kt}, as in \citet{Sussex2026-am}.

To pre-train cVS models, we first optimized $\theta$ under a forward KL objective.
To construct its training dataset, we drew 20 million samples from the prior and reweighted them by the reward, $\exp(\alpha r(x))$, picking a weight $\alpha$ such that the essential sample size was 1 million.
PGLD was pre-trained by sampling starting templates according to this tilted distribution, as described in~\citet{Sussex2026-am}.

\subsection{Details on TCRm design} \label{apx:scFv}
We used a transformer model trained on a dataset of scFv-pHLA interactions. Briefly, the data was generated by assembling a CDRH3 variational synthesis library (designed with EM VS) into second generation scFv-CAR constructs, delivering them into human cells, staining them with a panel of 100 fluorescently labeled and DNA-barcoded pHLA targets (dextramers), sorting for fluorescence, and single cell sequencing to read out the scFv sequence and counts of the number of dextramers bound to each cell \citep{Weinstein2026-vs,Unknown2026-og}.
Then, we trained an encoder-only transformer model with 7M parameters to predict binding counts from scFv CDRH3 sequence and pHLA, using LIFT and LeaVS \citep{Weinstein2025-fv,Weinstein2026-eg}.

To pre-train cVS with the forward KL objective, we drew 20 million samples from the prior (the initial variational synthesis model) and reweighted them by the reward, picking $\alpha$ such that the effective sample size was 5\% of the corpus, after dropping samples with an internal stop codon.
We pre-train to convergence, which took 8000 steps.
We then fine-tuned with the reverse KL objective for 5 million reward evaluations, with a batch size of 1024. We sweep 5 learning rates (0.003, 0.01, 0.03, 0.1, 0.3) and pick the one with the best final objective. We use cosine annealing with a 1000 step warmup, and anneal $\alpha$ from $2\alpha$ to $\alpha$ over the first half of training to encourage exploration (we found these annealing methods harmed PGLD, so only included them in cVS).

\subsection{Details on promoter engineering} \label{apx:promoter}

\begin{figure}
    \centering
    \includegraphics[width=\linewidth]{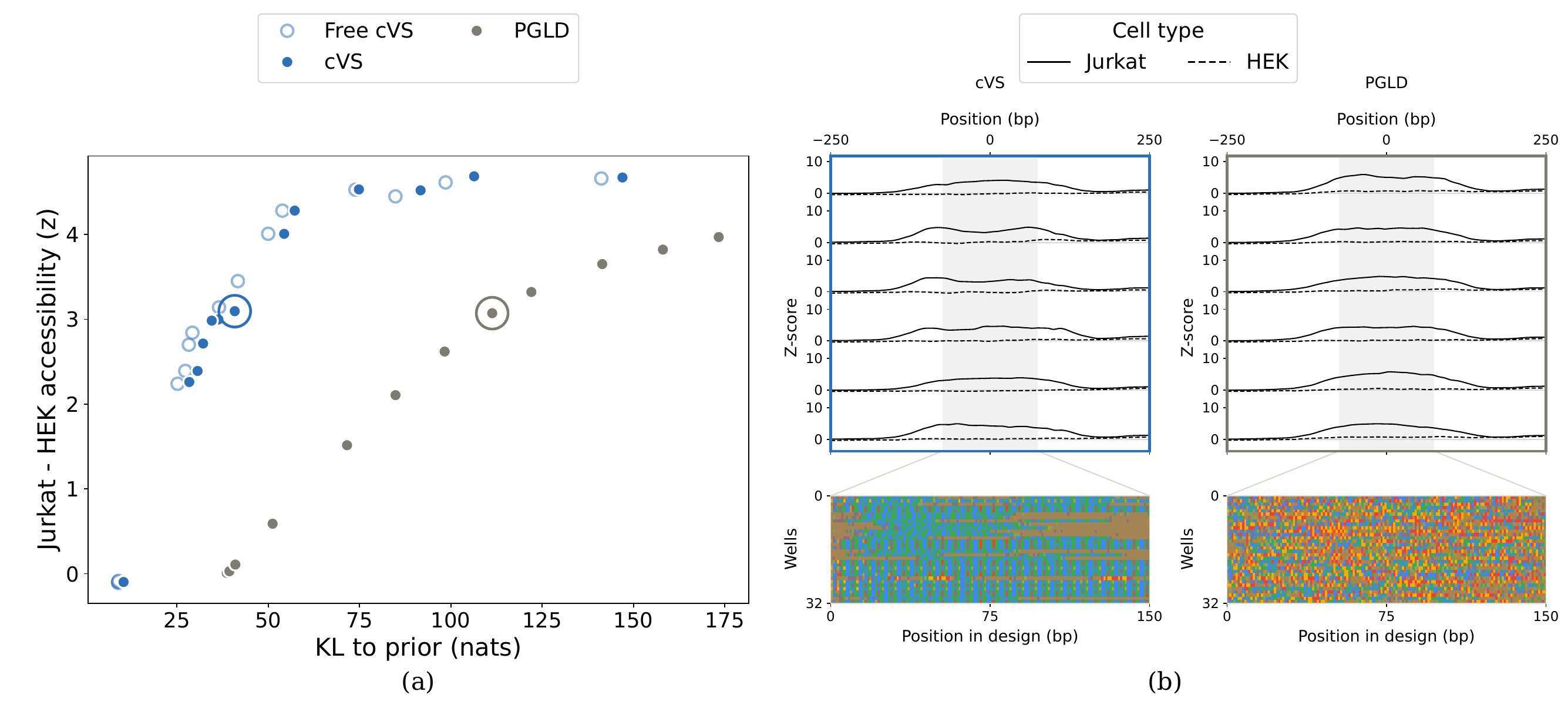}
    \caption{\textbf{Designed regulatory DNA sequences in a genomic context}. Same as \Cref{fig:atac_summary} but the designs are done in a random genomic context, rather than EF1$\alpha$.
    (a) Quality-diversity Pareto frontier, evaluating the average difference in cell-type accessibility versus the KL to the human genome prior. (b) Predicted accessibility of sampled sequences (above) from the learned synthesis models (below). Positions in each well are colored by the nucleotide mixture.}
    \label{fig:atac_summary_random}
\end{figure}

We downloaded ATAC seq data from ChIP-Atlas: SRX25532286, SRX7785407 and SRX16046833 for Jurkat, and SRX10665050, SRX7030829 and SRX3511089 for HEK \citep{Zou2024-eq}.
We evaluated accessibility in a custom plasmid context, where EF1$\alpha$ is used to drive expression of an scFv CAR. 
We designed the VS library into the first 129bp of EF1$\alpha$.
The genomic context we used for \Cref{fig:atac_summary_random} was chr10:63519858-63521858, which was near a peak in all 6 experiments above.
We design the centre 150 bp of the region.

BPnet was trained over all peaks in the 6 experiments above as well as an equal amount of GC-matched inter-genic regions as negatives.
Peaks from chromosomes 9, 10, 11 were held-out during training and used to early stop.
We trained 5 models at learning rates $0.001, 0.003, 0.01, 0.03, 0.1$ and picked the one with the best held-out loss.

MarinDNA was fine-tuned on the same peaks as BPnet.
Peaks from chromosomes 9, 10, 11 were held-out during training and used to early stop.
We trained 5 models at learning rates $0.001, 0.003, 0.01, 0.03, 0.1$ and picked the one with the best held-out loss.

We trained variational synthesis models with cVS and PGLD in the same way as in \Cref{apx:epitopes}, except tilting to an effective sample size of 0.5\% for pre-training.

\section{\textit{In vitro} evaluation} \label{apx:in_vitro}

We sampled 250ng of DNA from the synthesized library. The sequencing library was prepared using xGen\texttrademark ssDNA \& Low-Input DNA Library Preparation Kit (IDT) following manufacturer’s instructions with the exception that post-extension cleanup was performed using MinElute PCR Purification Kit (Qiagen). Dual indexed library samples were quantified using Kapa Library Quantification Kit (Roche). Pooled library samples were sequenced in a 2x150bp configuration on a NextSeq 2000 using P1 Reagents, 300 cycle Kit (Illumina). The resulting paired reads were basecalled with \texttt{bcl2fastq} v2.20 and assembled with \texttt{PEAR}. We obtained $24{,}471{,}996$ full-length reads.

 We used KSD-B (Kernelized Stein Discrepancy for Biological Sequences) to test how close the synthesis models are to the target $\pi(x) \propto p(x)\exp(\alpha r(x))$, where $p$ is the fine-tuned MarinDNA prior and $r$ the ATAC reward model. We used the implementation of \citet{Amin2023-dc}, with energy function $\log p(x) + \alpha r(x)$. Hyperparameters: an IMQ Hamming kernel ($c=1$, $\beta=0.5$), single-substitution neighbourhoods only; $m=5$ neighbour samples per point for the discrete Stein operator; and $n=1{,}000$ samples per model or from the sequencing data, redrawn 5 times with independent draws to report mean $\pm$ SEM. 

To compute the low-dimensional representations of the synthesis samples, we used the same procedure as in \cite{Weinstein2026-vs}: we first computed a matrix of pairwise Hamming distances between the samples from all evaluated models and the sequencing data, then transformed this distance matrix into Gram matrix using an IMQ Hamming kernel, and finally reduced the dimension of this matrix using UMAP \citep{mcinnes_umap_2020}.

\section{Supplemental Results}

\begin{figure}[h]
    \centering
    \includegraphics[width=\linewidth]{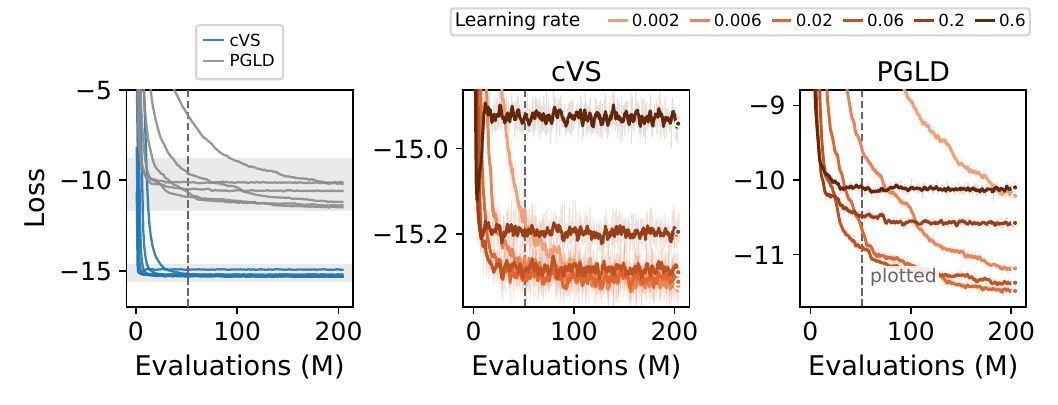}
    \caption{\textbf{Training loss as a function of the number of reward evaluations (in millions), for PGLD versus free cVS on the peptide benchmark}. Left: overview, with dashed line showing the results plotted in \Cref{fig:benchmark-summary}. Middle, Right: zoom in, with different learning rates marked.}
    \label{fig:benchmark_eval_scan}
\end{figure}

\begin{figure}
    \centering
    \includegraphics[width=\linewidth]{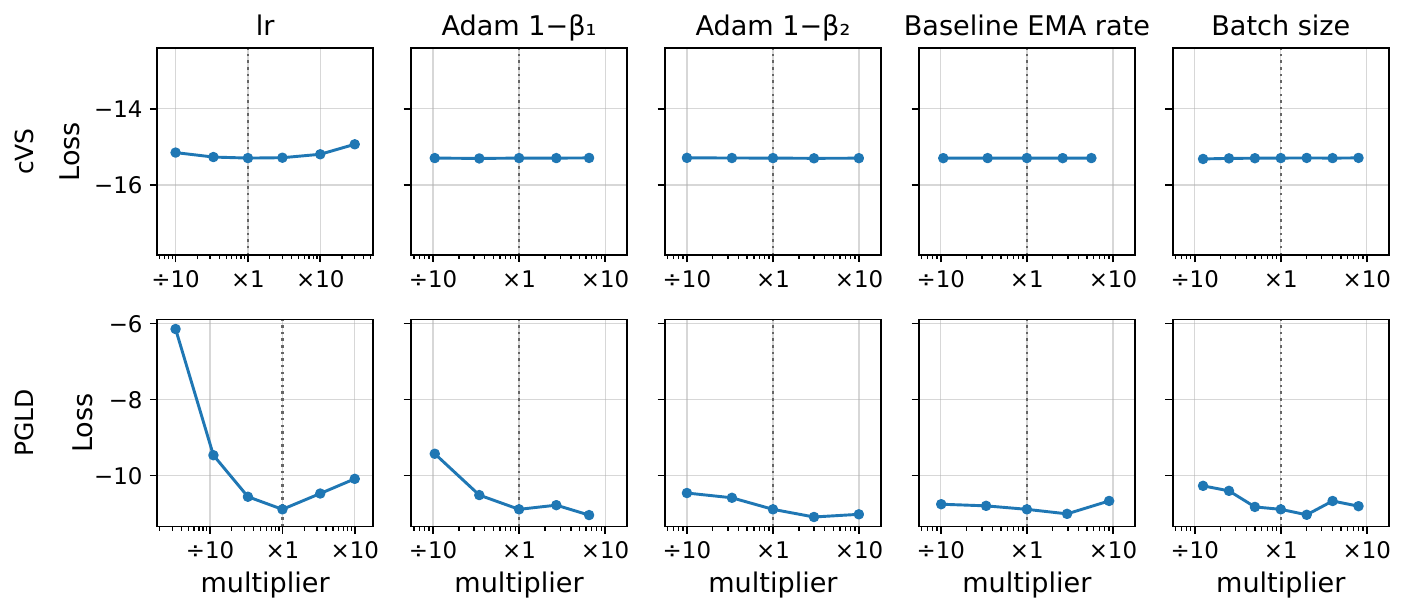}
    \caption{\textbf{Hyperparameter sensitivity analysis on the peptide benchmark, for PGLD versus free cVS}. x-axis scans each hyperparameter relative to the setting used in \Cref{fig:benchmark-summary}. First column: learning rate. Second and third columns: Adam hyperparameters. Fourth column: rate of the exponential moving average used to construct the baseline control variate. Fifth column: training batch size.}
    \label{fig:benchmark_hypers}
\end{figure}

\begin{figure}
    \centering
    \includegraphics[width=\linewidth]{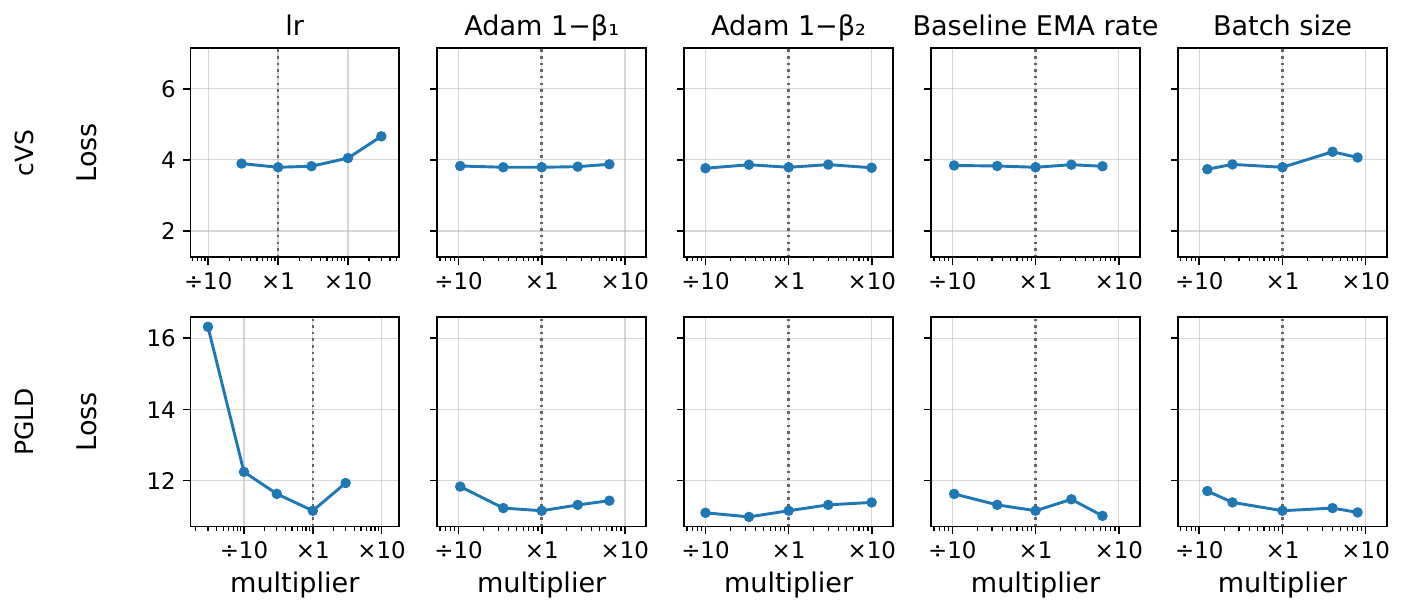}
    \caption{\textbf{Hyperparameter sensitivity analysis on scFv CDRH3 designs, for PGLD versus free cVS}. x-axis scans each hyperparameter relative to the setting used in \Cref{fig:mesa_summary}. First column: learning rate. Second and third columns: Adam hyperparameters. Fourth column: rate of the exponential moving average used to construct the baseline control variate. Fifth column: training batch size.}
    \label{fig:mesa_hypers}
\end{figure}

\begin{figure}
    \centering
    \includegraphics[width=\linewidth]{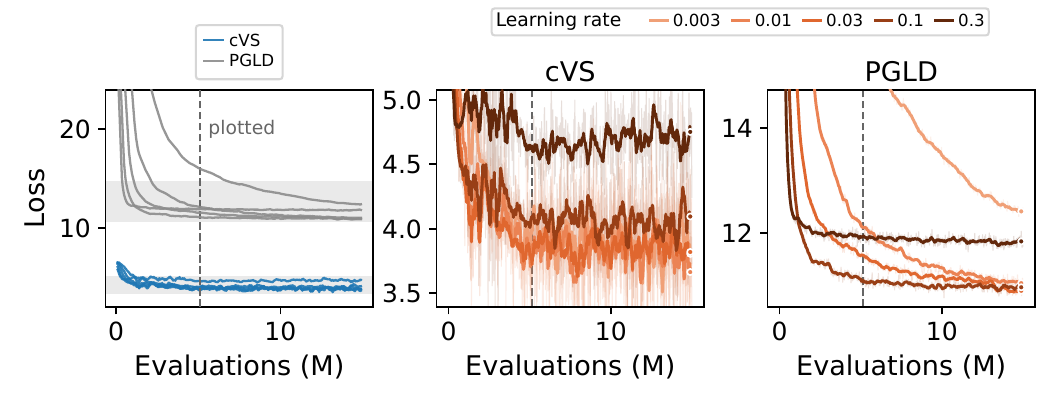}
    \caption{\textbf{Training loss as a function of the number of reward evaluations (in millions), for PGLD versus free cVS, on the scFv CDRH3 designs}. Left: overview, with dashed line showing the results plotted in \Cref{fig:mesa_summary}. Middle, Right: zoom in, with different learning rates marked.}
    \label{fig:mesa_eval_scan}
\end{figure}

\begin{figure}
    \centering
    \includegraphics[width=\linewidth]{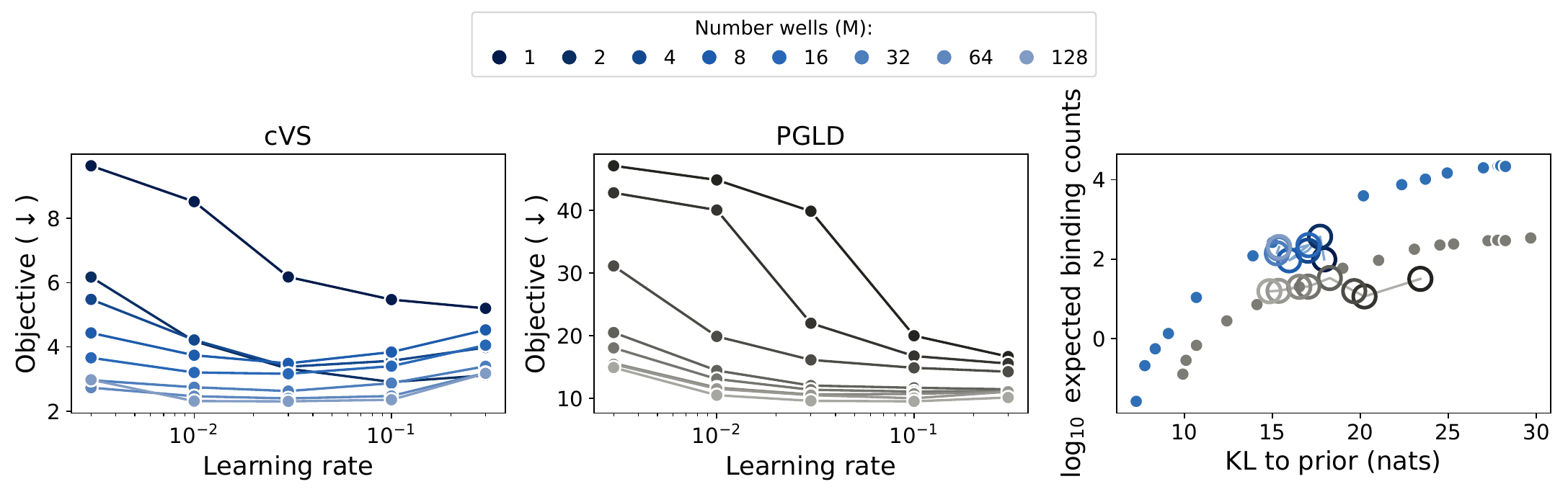}
    \caption{\textbf{Performance of cVS and PGLD as the number of wells $M$ increases, on the scFv CDRH3 designs.}
    Left: free cVS. Middle: PGLD. Right: post-quantization cVS for increasing $M$, compared to PGLD with increasing $M$.}
    \label{fig:mesa_M_scan}
\end{figure}

\begin{figure}
    \centering
    \includegraphics[width=0.5\linewidth]{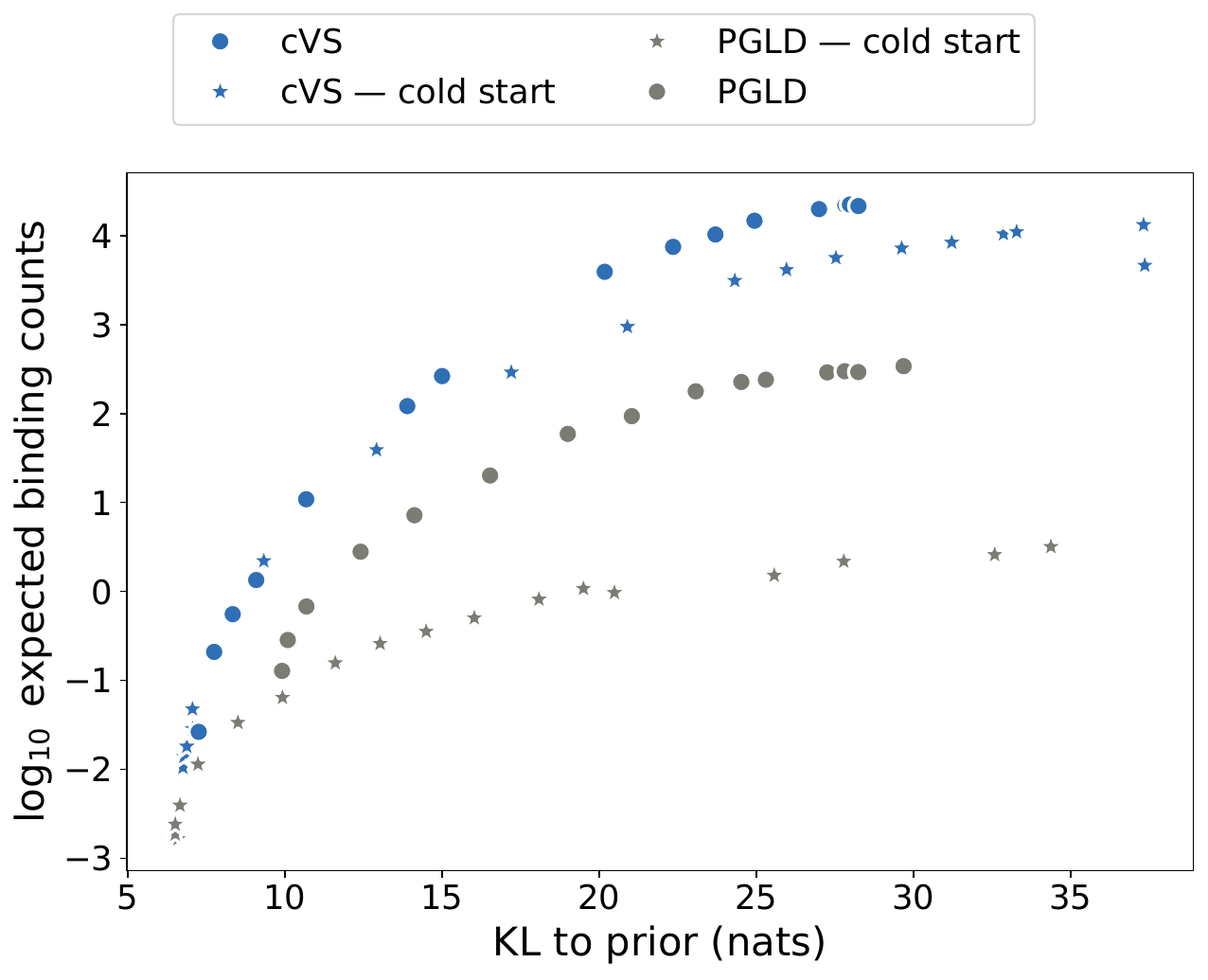}
    \caption{\textbf{Performance of cVS and PGLD with and without pretraining on the forward KL objective, on the scFv CDRH3 designs.}}
    \label{fig:mesa_pretrain}
\end{figure}

\begin{figure}
    \centering
    \includegraphics[width=0.5\linewidth]{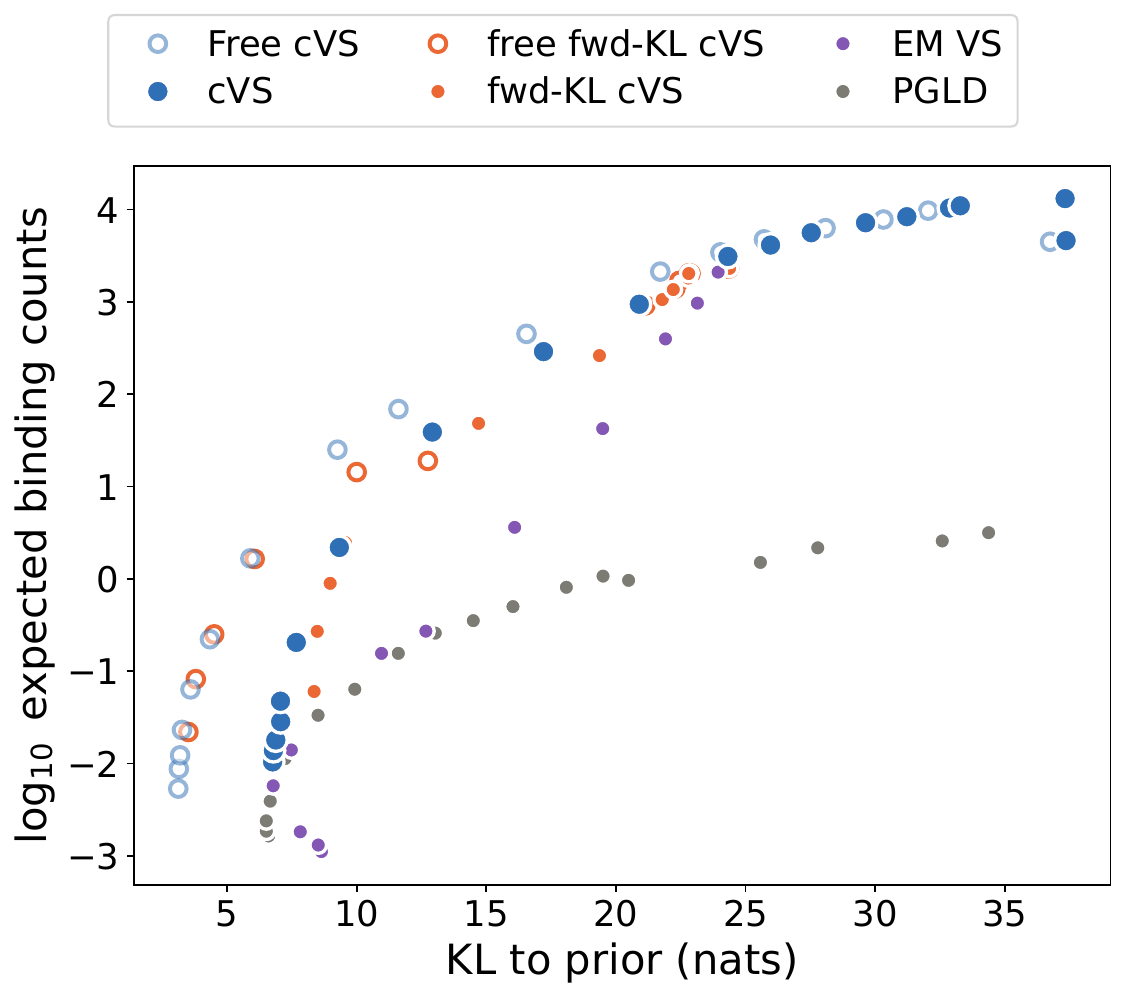}
    \caption{\textbf{Performance on the scFv CDRH3 designs with total reward evaluations held fixed.} Here each method is trained from scratch, using 5 million evaluations of the reward and prior, rather than including an added pre-training phase for cVS and PGLD as in \Cref{fig:mesa_summary}.}
    \label{fig:mesa_matched_eval}
\end{figure}

\begin{figure}
    \centering
    \includegraphics[width=0.5\linewidth]{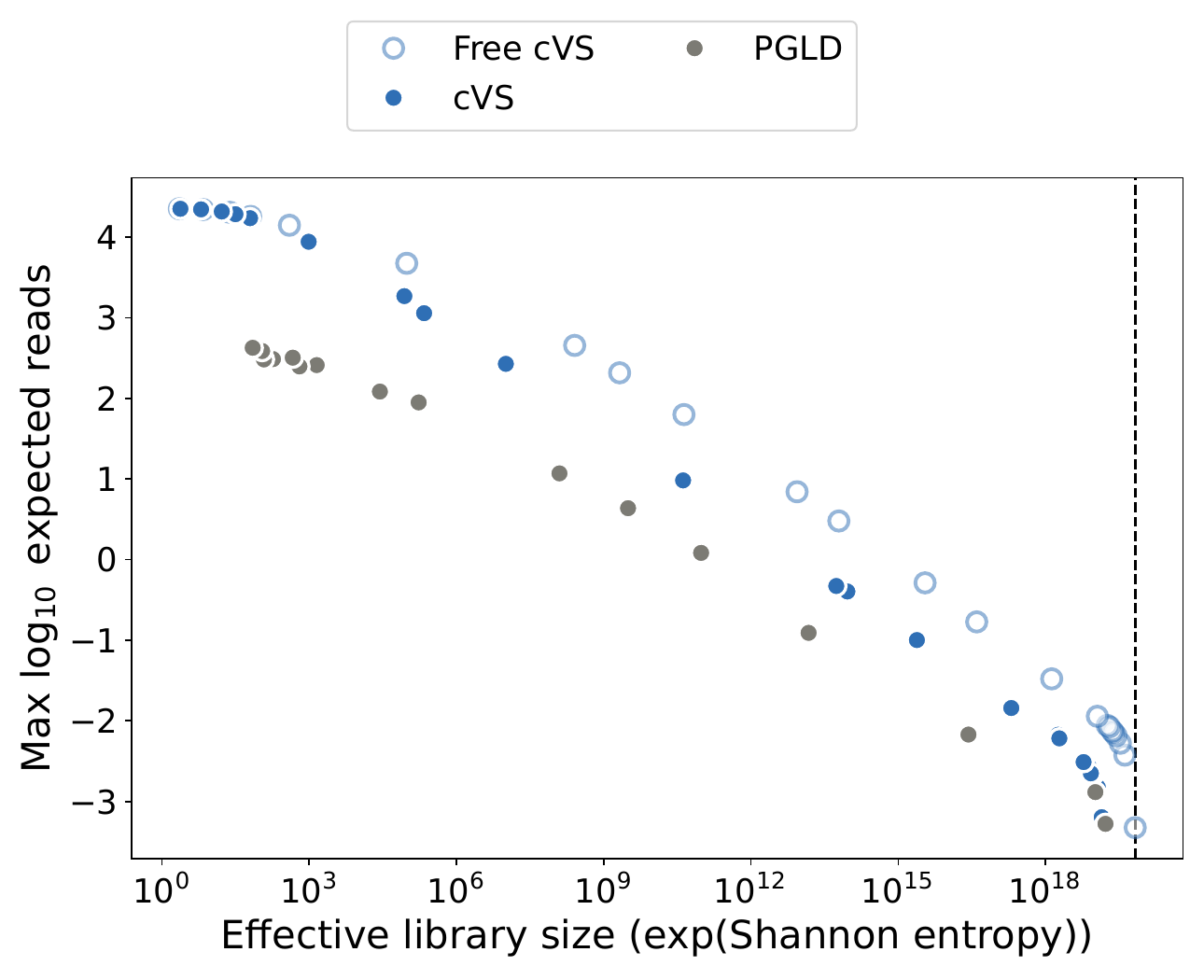}
    \caption{\textbf{Pareto frontier for cVS versus PGLD, measuring diversity by the Shannon entropy, on the scFv CDRH3 designs}. X-axis is the exponential of the Shannon entropy of $q_\theta(x)$, a measure of the effective library size. These designs are trained on a uniform prior rather than a human prior, to optimize for this diversity measure.}
    \label{fig:mesa_shannon_entropy}
\end{figure}

\begin{figure}
    \centering
    \includegraphics[width=0.7\linewidth]{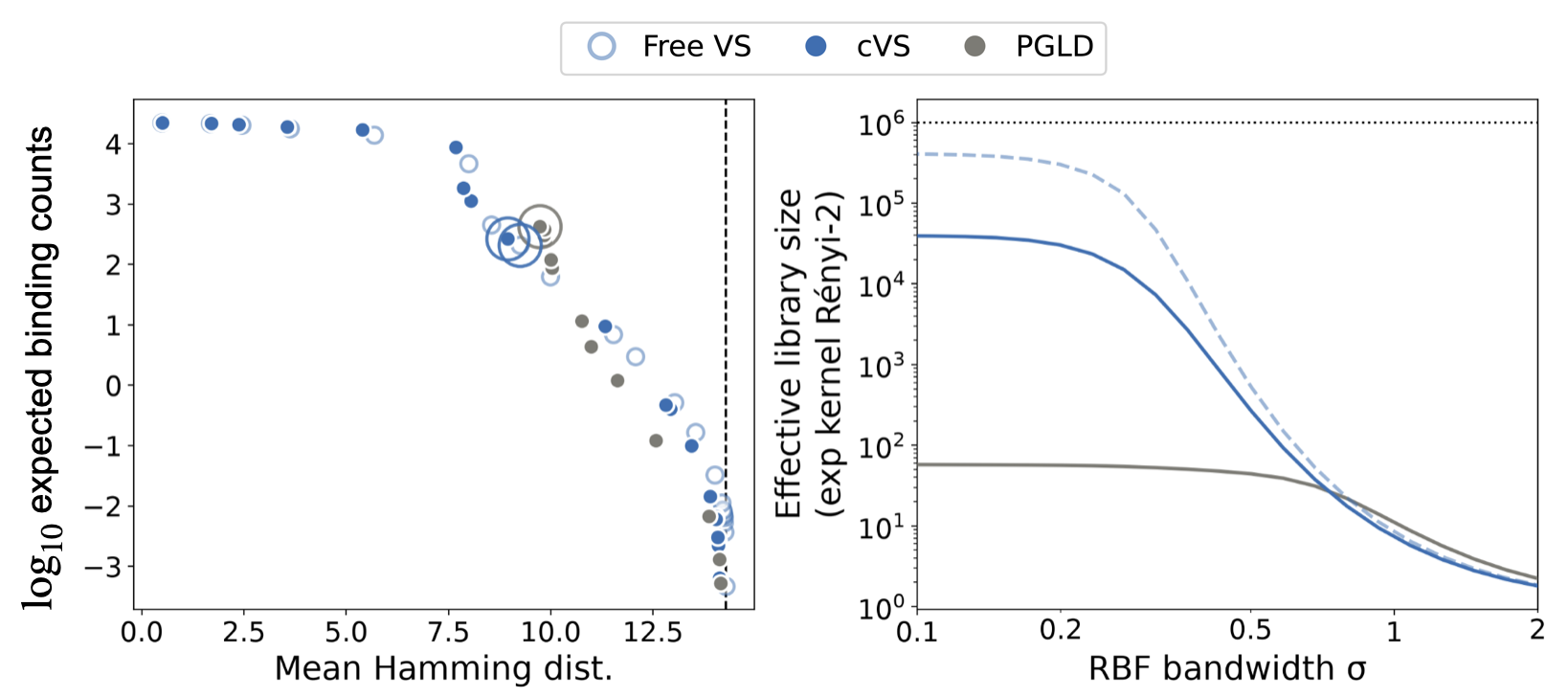}
    \caption{\textbf{Diversity measured by the exponential of the kernel 2-Renyi entropy, on the scFv CDRH3 designs.} We compare a cVS design to a PGLD design with the same constraints and same average reward, which achieves similar mean Hamming distance among sequences (left). We find similar diversity at high RBF kernel bandwidths, but cVS reaches much higher diversity at low bandwidths, indicating PGLD finds spread out but narrow modes compared to cVS (right).}
    \label{fig:mesa_renyi_entropy}
\end{figure}

\begin{figure}
    \centering
    \begin{subfigure}{0.582\textwidth}
    \includegraphics[width=\linewidth]{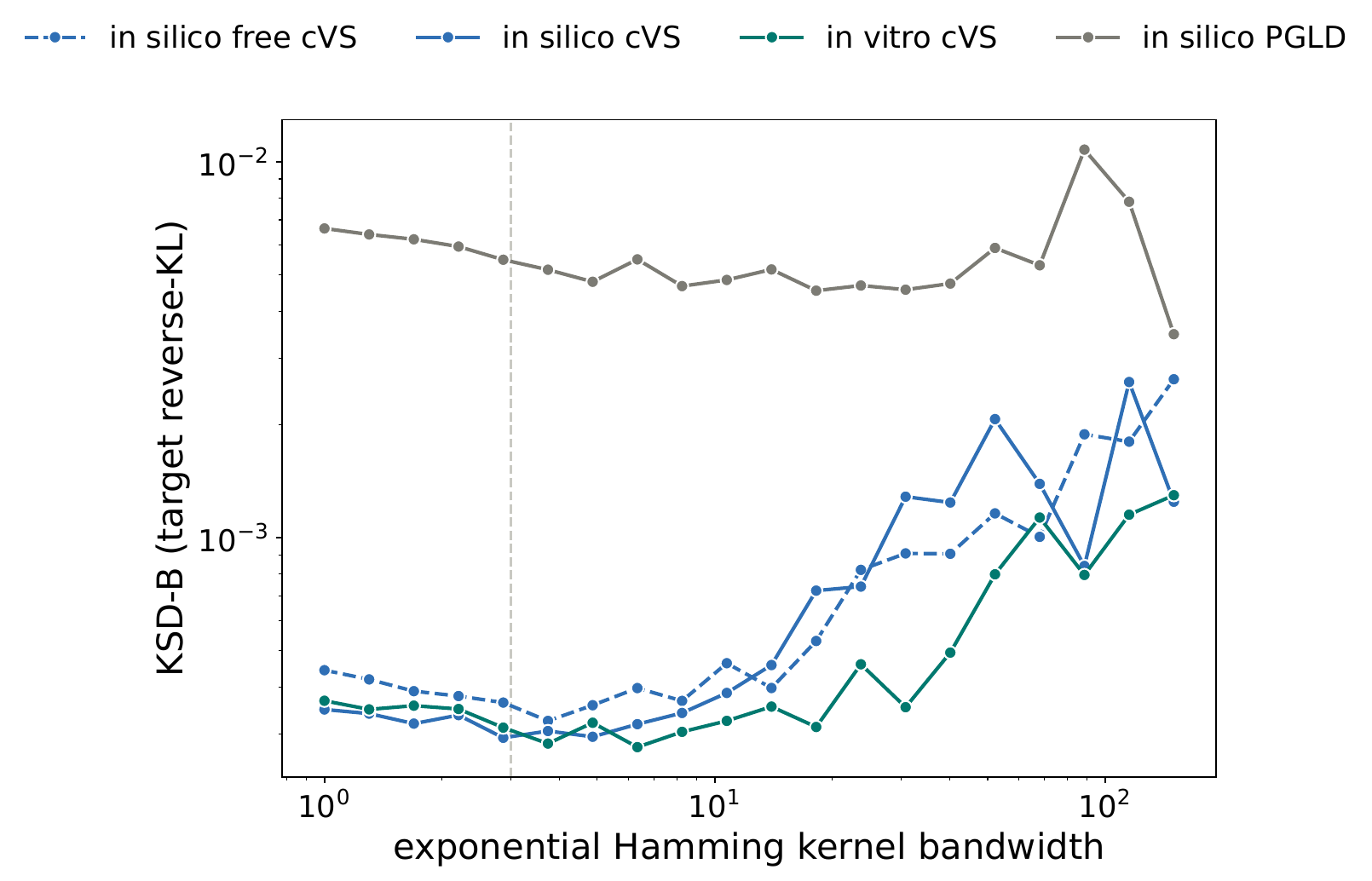}
    \caption{} \label{fig:ksdb_exp_sweep}
    \end{subfigure}
    \begin{subfigure}{0.3\textwidth}
    \includegraphics[width=\linewidth]{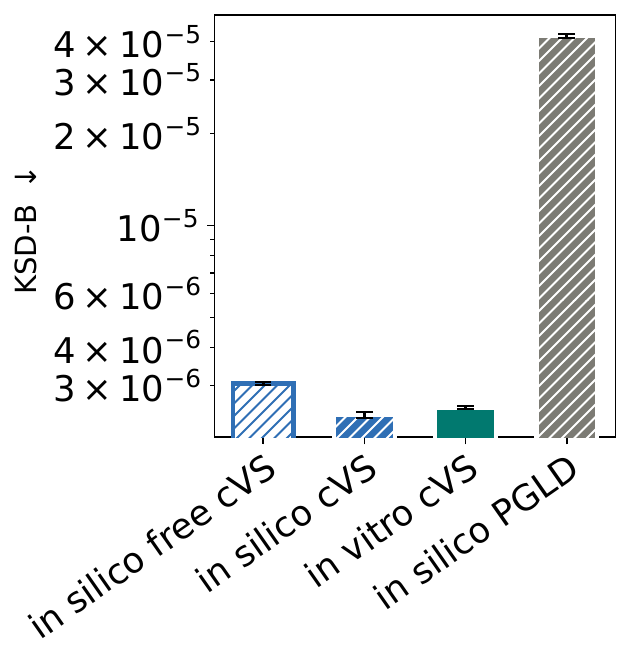}
    \caption{}\label{fig:ksdb_exp_bar}
    \end{subfigure}
    \caption{\textbf{KSD-B of EF1$\alpha$ libraries computed with with exponential Hamming kernels}
    (a) KSD-B as in \Cref{fig:atac_in_vitro_ksdb} but computed with an exponential Hamming kernel with a scanned bandwidth $\sigma$, rather than an IMQ Hamming.
    (b) Same as in (a) but with a fixed $\sigma=3$ and five independent redraws for each model (error bars: SEM across redraws)
    }
    \label{fig:atac_in_vitro_ksd_sensitivity}
\end{figure}

\end{document}